%% file: main.tex
\documentclass[copyright,creativecommons]{eptcs}
\usepackage{tikz}
\usetikzlibrary{shapes,arrows}
\usepackage{graphicx}

\usepackage{url}
\usepackage{paralist}
\usepackage{xcolor}
\usepackage[english,noabbrev]{cleveref}

\usetikzlibrary{positioning,fit,arrows.meta,backgrounds}
\tikzset{
    module/.style={%
        draw, rounded corners,
        minimum width=#1,
        minimum height=7mm,
        font=\sffamily
        },
    module/.default=2cm
}
\definecolor{background}{HTML}{282a36}
\definecolor{foreground}{HTML}{44475a}%6272a4
\definecolor{greynode}{HTML}{6272a4}%44475a
\definecolor{pinks}{HTML}{ff79c6}%282a36 %ff79c6
\definecolor{oranges}{HTML}{ffb86c}
\definecolor{cyans}{HTML}{8be9fd}
\definecolor{feedback}{HTML}{171920}

\usepackage{xspace}
\newcommand{\ProofBuddy}{\textsc{ProofBuddy}\xspace}
\newcommand{\ProofBuddyv}[1]{\textsc{ProofBuddy v#1}\xspace}

\ifpdf
  \usepackage{underscore}         % Only needed if you use pdflatex.
  \usepackage[T1]{fontenc}        % Recommended with pdflatex
\else
  \usepackage{breakurl}           % Not needed if you use pdflatex only.
\fi

\title{\ProofBuddy\thanks{
    Our tool \ProofBuddy should not be confused with another tool called “Proof Buddy”~\cite{Earth2023}.
    We first presented an extended abstract about \ProofBuddy~\cite{karsten.etal-TFPIE23} in January 2023 at the TFPIE Workshop~\cite{TFPIE2023}, published as a more comprehensive description in the post-proceedings in August 2023~\cite{proofbuddy2023}. “Proof Buddy” is a tool by Steve Earth, Jeremy Johnson and Bruce Char, presented at a demo session in March 2023 at SIGCSE 2023~\cite{Earth2023}.}\\--- How it Started, How it's Going ---}

\author{
  Nadine Karsten
  \institute{Technische Universität Berlin, Germany}
  \email{n.karsten@tu-berlin.de}
  \and
  Kim Jana Eiken
  \institute{Technische Universität Berlin, Germany}
  \email{k.eiken@campus.tu-berlin.de}
  \and
  Uwe Nestmann
  \institute{Technische Universität Berlin, Germany}
  \email{uwe.nestmann@tu-berlin.de}
}

\newcommand{\titlerunning}{\ProofBuddy}
\newcommand{\authorrunning}{N.\ Karsten, K.\ J.\ Eiken \& U.\ Nestmann}

\hypersetup{
  bookmarksnumbered,
  pdftitle    = {\titlerunning},
  pdfauthor   = {\authorrunning},
  pdfsubject  = {EPTCS},               % Consider adding a more appropriate subject or description
}

\begin{document}

\maketitle

\begin{abstract}
  We report on our journey to develop \ProofBuddy, a web application that is powered by a server-side instance of the proof assistant Isabelle, for the teaching and learning of proofs and proving.
  The journey started from an attempt to use just Isabelle in an educational context.
  Along the way, following the educational design research approach with a series of experiments and their evaluations, we observed that a web application like \ProofBuddy has many advantages over a desktop application, for developers and teachers as well as for students.
  In summary, the advantages cover simplicity, maintainability and customizability.
  We particularly highlight the latter by exhibiting the potential of interactive tutorials and their implementation within \ProofBuddy.
\end{abstract}

% The table of contents below is added for your convenience. Please do not use
% the table of contents if you are preparing your paper for publication in the
% EPiC Series or Kalpa Publications series

\setcounter{tocdepth}{2}

%\section{To mention}
%
%Processing in EasyChair - number of pages.
%
%Examples of how EasyChair processes papers. Caveats (replacement of EC
%class, errors).

%------------------------------------------------------------------------------

\input{introduction}
\input{proof-assistants}
\input{proofbuddy}
\input{interactive-tutorials}
\input{conclusion}

%------------------------------------------------------------------------------

\newpage

\label{sect:bib}
\bibliographystyle{eptcs}
\bibliography{references}

\newpage
\appendix
%{\small \tableofcontents}

%------------------------------------------------------------------------------

\end{document}

%% file: introduction.tex
\section{Introduction \label{sec:introduction}}

\paragraph{Motivation.}

Students in Computer Science (CS) often encounter difficulties in developing and writing proofs by themselves, as highlighted, e.g., by Frede and Knobelsdorf~\cite{FredeKnobelsdorf2018,analysingStudentPracKnobelsdorf16}.
The primary challenges lie in the use of ``formal'' and ``mathematical'' language, but also in assessing the correctness and completeness of proofs~\cite{Kiehn2017}.
As CS students are used to interact with computers for acquiring competences in programming, among other activities, it seems natural to provide CS students also with computer-based tools for the acquisition of competences in proving.
Various strategies have been proposed along these lines aiming to provide students with much more immediate and personalized feedback on their own learning process.

In the remainder of this paper, we avoid the use of the terms ``formal'' and ``mathematical'' language, as they can be misleading.
Instead, we stick to the following convention for three levels of language with varying degrees of ``formality''~\cite{Boehne2019}:
\begin{inparaenum}[(i)]
\item We use \emph{natural language} to refer to what one can find in typical textbooks, i.e., a ``pseudo-natural'' language, according to socio-mathematical norms~\cite{tatsisa08}, sprinkled with mathematical or logical formulae.
\item We use \emph{logical language} to refer to a formal representation of proofs within some logical system, e.g., using natural deduction~\cite{gentzen:untersuchungen1,fitch1952symbolic}.
\item We use \emph{tool [input] language} to refer to the input language of a tool (\Cref{sec:proof-assistants}) that supports the construction of proofs.
\end{inparaenum}
\newpage
The ``teaching and learning of proofs and proving'' (TLPP) is of course \emph{not} a topic originating \emph{only} from CS, but has a long tradition in the field of Mathematics education.
There, the notion of \emph{proof competence} has been developed and refined over many years.
In order to use this notion to classify and evaluate the possible effects of computer-based tools for TLPP, we explain it in more detail.
Proof competence typically includes four dimensions of competence~\cite{Brunner2014,krieger2015mathematische}:
(1)~\emph{Professional competence} comprises the knowledge about the topic of the proof task.
(2)~\emph{Representation competence} describes the ability to use a ``sufficiently formal'' language to write a proof.
(3)~\emph{Communication competence} characterizes the ability to discuss the finding of a proof and the solution.
(4)~\emph{Methodological competence} comprises three parts \cite{Heinze2003,Brunner2014}: (i)~the \emph{proof scheme} involves the use of deductive patterns for proving assertions, (ii)~the \emph{proof structure} entails the systematic organization of a proof and (iii)~the \emph{chain of conclusions} refers to the sequential arrangement of steps within a proof, forming a coherent chain of arguments.

\paragraph{\ProofBuddy.}

In~\cite{proofbuddy2023}, we introduced our own computer-based tool to support TLPP:
\ProofBuddy\footnote{\url{proofbuddy.tu-berlin.de}} is a web application that is based on the interactive theorem prover Isabelle, having it running in the backend.
Opting for a web application format as opposed to the use of Isabelle/jEdit~\cite{IsabellejEdit} (the IDE that is part of the Isabelle software package) has the advantage of enabling the collection of a rich dataset for analyzing learning behaviors and outcomes, next to the unneeded local installation of Isabelle~\cite{proofbuddy2023}.

As also announced in~\cite{proofbuddy2023}, we carried out teaching experiments via a dedicated TLPP-course at Bachelor level to improve and evaluate our tool \emph{and} our TLPP-materials.
We decided to follow the \emph{educational design research} approach (see
next paragraph) to systematically and consciously drive the development
process of \ProofBuddy from \textsc{v0} up to \textsc{v2}.

\paragraph{Educational design research.}

Educational design research (EDR) \cite{nieveen2006,mckenney2018,plomp2013} refers to a comprehensive approach in which educational theories are applied to concrete educational challenges, for example, to improve a university course to better achieve its goals.
In this context, researchers and teachers often work closely together to try out creative ideas and to test them continuously and iteratively.
% including intervention, iteration, and a focus on process, utility, and theory.
First, the given educational challenge must be carefully analyzed to work out suggestions for specific interventions.
Then, researchers design systems, often using technologies, to devise teaching and learning materials and methods to achieve the predicted learning gains based on some educational theory.
The response of teachers and students to specific design features suggested by the theory is then examined by the designers as the systems are put into practice and analyzed.
The researchers then start the next iteration to either try out another idea for the original problem or to identify the next biggest problem.
Given the inherent uncertainties of this process, numerous smaller and less controlled studies about EDR have been conducted. %, each addressing a single question.
In a given concrete EDR scenario, it is important to focus on one problem at a time, to understand it, to determine the interventions that might lead to success, to build prototypes and to test them.
As a matter of fact, EDR carries the risk of not necessarily leading to clearly identifiable or measurable success.
While the first iteration often leads to weak results, the availability of various design options facilitates the refinement and enhancement of outcomes through direct comparisons across iterations.
The results of such studies are meaningful even if there is no control group (which would be useful to compare the outcome of an experiment with and without a tested intervention), because the changes between the iterations are usually small.

\paragraph{Contribution.}

In this paper, we report on our journey to iteratively run teaching experiments following the EDR approach and share our learnings along the way.
We developed \ProofBuddy over three iterations, starting with the prototype \ProofBuddyv{0}, as presented in \cite{proofbuddy2023}, in order to facilitate TLPP.
To better guide students on their learning path, we introduce interactive tutorials for \ProofBuddyv{1}. To the best of our knowledge, the iterative and \emph{joint} development of a TLPP-tool, TLPP-materials and an associated course structure, \emph{together} with systematic evaluations that are rooted in didactics research and that are based on increasingly rich datasets is novel in this context.
One goal of this paper is to have the readers understand the various stages of the development---up to now, but also for the next steps, including \ProofBuddyv{2}---and, at best, to agree with our findings and conclusions.

\paragraph{Overview.}

In Section~\ref{sec:proof-assistants}, we explain proof assistants in some generality---with a focus on how they can be used in teaching.
%The possibility to collect data motivated us to develop the first version of the web application \ProofBuddy\footnote{\url{proofbuddy.tu-berlin.de}} \cite{proofbuddy2023} to facilitate the introduction of the basics of proof structures and proof techniques across various levels of expertise.
%\KE{Gefällt mir noch nicht}
%\UN{Könntest Du das konkretisieren?}
%\UN{Mir fällt gerade auf, dass das eigentlich zum nächsten Paragraphen gehört.}

In Section~\ref{sec:proofbuddy}, we present \ProofBuddy including an overview of its software architecture.
Furthermore, we recapitulate a usability study of the prototype \ProofBuddyv{0} conducted in \cite{proofbuddy2023}, including an evaluation of the results.
One outcome of the usability study was that the TLPP with \ProofBuddyv{0} would profit from a much more guided style in the sense of (interactive) tutorials (on which we report in Section~\ref{sec:inter-tutor}).

In \Cref{sec:eduDeI}, we describe the first iteration of our test course.
To enable a direct comparison between Isabelle/jEdit and \ProofBuddy and because the infrastructure was not yet fully operational at this time, we used the Isabelle/jEdit instead of \ProofBuddyv{0} for this iteration.
During the course, we conduct our first experiment in a somewhat limited form and analyze its outcome. The findings suggest that while a PA's immediate feedback can support students in learning how to prove, the complexity has to be reduced.

In \Cref{sec:inter-tutor}, we focus on the structure and composition of interactive tutorials for PAs.
We highlight how such tutorials may support TLPP, report on related work and, finally, introduce interactive tutorials in \ProofBuddy.

In \Cref{sec:eduDeII}, we report on our second course iteration using \ProofBuddyv{1}, together with its evaluation, this time with an expanded data collection by recording the teacher’s voice as well as the student’s questions.
%supported by substantial data collection \UN{of what kind?}.
%Compared to the first iteration (see Subsection 4.1), we spent considerably more time on natural deduction and introduced it—being a logical language—as a “bridge” between natural language and the Isar language.
%The collection of questions that students asked during the course sessions was expanded. We additionally
%documented where the students pointed at, what the teacher’s answer was and what medium
%was chosen for the explanation. To do this, we recorded the teacher’s voice as well as the
%student’s questions

\Cref{sec:conclusion} concludes the paper on the basis of two EDR-iterations.

In \Cref{sec:futureWork}, we briefly report on our current experiment with a further improved version of \ProofBuddyv{2} and further developed tutorials.
Additionally, we shortly introduce our plans for a forth EDR-iteration, the usage of \ProofBuddy in a compulsory CS Bachelor course and a tutorial management system for \ProofBuddy with integrated learning analytics.

The journey continues \ldots

%%% Local Variables:
%%% mode: latex
%%% TeX-master: "main"
%%% End:

%% file: proof-assistants.tex
\section{Proof Assistants}
\label{sec:proof-assistants}

Interactive theorem provers, also referred to as proof assistants (PAs), are software tools that are engineered to allow the user to develop proofs step-by-step using a dedicated tool input language, which we therefore also refer to as PA language.
PAs are interactive, as they offer immediate feedback to the user on the accuracy of individual proof steps and the overall completeness of proofs.
Each component of a proof must adhere to logical rules, with every step being rigorously verified to ensure the absence of gaps in proof chains.
%Deviation from logical principles is not permitted, so that a valid proof is provided that is thoroughly checked.
% at every proof state.\NK{I do not like this sentence. ProofState has a special meaning in the Isabelle context.}
% PAs require the use of a formal language and scrutinize each step of a proof.
%

\subsection{Spoilt for Choice}\label{sec:weapon}
Nowadays, the probably most popular PAs are Coq \cite{bookCoq}, Lean \cite{Lean} and Isabelle \cite{Nipkow-Paulson-Wenzel:2002,IsabelleProving,wenzel2014isabelle}.
They are all based on some underlying logic, a kernel consisting of axioms and interference rules, that allows to formally deduce, i.e.\ prove, further logical statements.
To avoid having to prove every single deduction step explicitly, PAs often include so-called tactics that automatically apply a combination of rules in one go.
Moreover, PAs like Isabelle also allow the usage of some external automatic theorem provers (ATP) that are more powerful than mere tactics.

Coq implements dependently typed functional programming, so types usually must be written explicitly.
Within the Coq language, rules and tactics are applied to the goal that generate a backward proof by manipulating the goal.
The Coq language for writing proofs is then just a script composed of rules and tactics.
Lean is a functional programming language with inductive types, where the types usually have to be written explicitly.
The Lean language allows, besides using scripts similar to those of Coq, to write proofs roughly resembling paper proofs in ``natural'' language, with a set of fixed keywords to structure proofs and the possibility to explicitly formulate assumptions, intermediate steps and goals, which makes proofs more readable for humans.
Isabelle, also based on typed functional programming, is a generic PA, in which concrete logics can be embedded.
But in contrast to Coq and Lean, it provides more automatic type inference when it comes to the basics of proving, i.e. propositional and first order logic.
In Isabelle, the extension with the Isar language \cite{isabelleIsar} is standard.
The Isar language allows to write structured PA proofs resembling natural language proofs nearly as in Lean.

Several studies, such as~\cite{inp:BoehneKnobelsdorfKreitz16a,coqToText}, indicate that the mere use of Coq does not significantly enhance proof writing skills.
This aspect and the fact that types usually have to be given explicitly led us to decide not to use this PA.
Nowadays there exist some language extensions, i.e. Waterproof \cite{wemmenhove2023waterproof}, which are optimize for Mathematical proofs. 
As we want to focus on the proofs themselves and to explain types only later in the course, we also decided not to use Lean.
Another fact is that Lean was not stable, i.e. the versions differs to much.
LeanVerbose\footnote{\url{https://github.com/PatrickMassot/lean-verbose}} also has this problem when upgrading from Lean3 to Lean4.
For \ProofBuddy we choose to base on Isabelle which offers the possibility of type inference and the Isar language~\cite{isabelleIsar}.
Additionally, the pedagogical experiences with Isabelle have been favorable in \cite{LSDproofsSemantic,Nipkow-CPP21,compuMethaphysics}, observing significantly improved structure in student-authored post-instruction proofs.
Since Isabelle was developed for experts and not for teaching, it is important to integrate Isabelle into teaching in a well-considered and adapted way.

\subsection{Teaching with Proof Assistants}
\label{sec:teaching-with-proof}

Like all learners or novices, also students require feedback on their own proofs.
Without software support, teachers or teaching assistants act as human proof assistants; they explain the finer points of proof construction and correctness during exercises and provide feedback on submitted homework.
However, feedback is often provided only days or weeks after homework submission.
Moreover, students are unable to make corrections and resubmit their work, so their opportunity and ability to improve proof competence (\Cref{sec:introduction}) is limited~\cite{Heinze2003,krieger2015mathematische}.

To address this problem, various approaches have been proposed to integrate PAs into teaching settings, enabling direct and individualized feedback to significantly improve the TLPP.
Some approaches \cite{LSDproofsSemantic,compuMethaphysics,KnobKreitz2017,coqToText} assume an experienced level of proof competence, while others~\cite{Avigad2019,wemmenhove2022waterproof} use PAs in introductory courses for first-year students to facilitate the learning process.
The idea of training especially the \emph{methodological} competence (\Cref{sec:introduction}) with the help of some PA language is not new and apparently leads to better structured proofs \cite{LSDproofsSemantic}.

Writing a proof in a PA feels like programming, so most CS students find it natural and even comfortable.
PAs typically display all remaining sub-goals; this helps students to decide what they should prove next.
The direct feedback also invites students to just try out proof techniques, to recognize when they get stuck, and enables them to correct themselves during the process of proving.
All of these mentioned points lead to some kind of gamification \cite{LSDproofsSemantic}, so students often have more fun with proofs, and also with the more theoretical part of CS.

Next to all the advantages of the use of PAs for TLPP, there are some disadvantages.
First, PAs demand from their users a significant level of expertise.
The complex syntax of PAs results in a steep learning curve, consuming substantial amounts of time for students.
Students are easily overwhelmed by the syntax and the extensive possibilities offered by PAs, even with just the interfaces of desktop or web applications of common PAs, which were simply not developed for teaching purposes.
Furthermore, proofs and feedback generated by PAs may not always be understandable for beginners, which hinders the learning process.
Moreover, the reliance on tactics and ATPs may impede students' understanding of proofs because they are powerful and convenient and may keep novices from actually learning the basics of logic.
Therefore, this support that is rather intended for expert users should be hidden from learners.

% Another issue is accessibility: students and teachers can use a local application that requires external distribution of the exercises, while when using a web application, data protection and the ability to save files must be guaranteed.

The use of PAs for TLPP requires that all students get access to the software, and ideally the same version of it.
There exist desktop and web applications for various PAs.
By nature, the former require local installations, while the latter just require access to the internet.
Both have advantages and disadvantages, which we discuss to some extent later on (e.g., in \Cref{sec:inter-tutor}).

We firmly believe that PAs can enhance---due to the better structured proofs---the overall proof competence of CS students, especially representation competence and methodological competence (see also~\cite{LSDproofsSemantic}).
PAs support the understanding of proof schemata via inference rules: proof steps in PAs are justified when all necessary assumptions of the associated rule are satisfied.
The quick feedback on the correctness of each step and on what remains to be shown in a goal gives users the opportunity to correct their mistakes; likewise, when building chains of reasoning in their proofs, users are provided with feedback on the overall structure and completeness of proofs.

%%% Local Variables:
%%% mode: latex
%%% TeX-master: "main"
%%% End:

%% file: proofbuddy.tex
\section{\ProofBuddy}
\label{sec:proofbuddy}

\ProofBuddy\footnote{\url{proofbuddy.tu-berlin.de}} is a web application for TLPP.
It is designed to enhance the process of learning, writing, and verifying proofs using the Isabelle proof assistant.
While the integrated development environment (IDE) for Isabelle, Isabelle/jEdit \cite{IsabellejEdit}, is a powerful application for proof construction, it is not suitable for undergraduate students due to its advanced features (Subsection~\ref{sec:evaluation-1}), giving direct access to the complex syntax of Isabelle, and causing the steep learning curve associated with it \cite{proofbuddy2023}.
\ProofBuddy, on the other hand, provides a gentle interface for Isabelle, offering a simplified user interface (UI), restricted syntax and explanations tailored to the subject at hand.
It is intended for TLPP in educational contexts (Subsection~\ref{sec:teach-with-proofb}), offering centralized course management and data collection features.

We developed the UI, features and infrastructure of \ProofBuddy iteratively, and continue to do so: We started with a prototype, \ProofBuddyv{0} (\Cref{fig:pb-v0}), in a usability study (Subsection~\ref{sec:teach-with-proofb}).
Based on the results we improved it and added interactive tutorials (\Cref{sec:inter-tutor}), \ProofBuddyv{1} (\Cref{fig:pb-v1}), used in the second iteration of WoP (\Cref{sec:eduDeII}).
The latest version, \ProofBuddyv{2} (\Cref{fig:pb-v2}), is currently used in the third iteration of WoP (\Cref{{sec:wop3}}).
\nopagebreak
\ProofBuddy's architecture has remained largely the same and is described below.

\subsection{Architecture}
\label{sec:architecture}

\begin{figure}
\centering
\begin{tikzpicture}[
            module/.style={rectangle, draw=foreground, fill=greynode, very thick, minimum width=2cm, minimum height=1cm},
            background-node/.style={module, draw=foreground},
            background-node2/.style={module, draw=foreground, minimum width=8cm, minimum height=5.5cm},
        ]

        % Nodes
        \node[module, dashed] (client) {web client};
        \node[module, dashed, below=2.3cm of client] (client2) {web client};
        \node[module, above=0.2cm of client, right=2cm of client] (server) {web server};
        \node[module, below=1.5cm of server] (database) {database};
        \node[module, dashed, right=1.5cm of database] (isabelle) {Isabelle};
        \node[module, dashed, above=.3cm of isabelle] (isabelle1) {Isabelle};

        % Labels
        \node[above=0.3cm of client, color=foreground] (browser) {\textbf{browser}};
        \node[above=0.3cm of client2, color=foreground] (browser2) {\textbf{browser}};
        \node[above=1.5cm of isabelle, color=foreground] (ism) {\textbf{ISM}};
        \node[above=of ism, color=foreground] (backend) {};%\textbf{Backend}

        % Background
        \begin{scope}[on background layer]
            \node[fit=(server) (database) (isabelle), background-node2, fill=greynode!50] {};
            \node[fit=(isabelle) (ism), background-node, fill=greynode!30] (ism) {};
            \node[fit=(client) (browser), dashed, background-node, fill=greynode!30] {};
            \node[fit=(client2) (browser2), dashed, background-node, fill=greynode!30] {};
        \end{scope}

        % Arrows
        \draw[<->, color=foreground, thick] (client2) -- (server);
        \draw[<->, color=foreground, thick] (client) -- (server);
        \draw[<->, color=foreground, thick] (server) -- (ism);
        \draw[<->, color=foreground, thick] (server) -- node[below] {} (database);
        \end{tikzpicture}
        \caption{Architecture of \ProofBuddy}
        \label{fig:architecture}
\end{figure}

\ProofBuddy's architecture comprises a web client, a web server, a database and the Isabelle Server Manager (ISM), as depicted in Figure~\ref{fig:architecture}.
The \emph{web client} is responsible for rendering the user interface.
In Isabelle, formulas can occur in a proposition and intermediate steps which are surrounded by keywords based on a given grammar separating formulas from rules and tactics.
The layer of the formulas is called inner syntax and the grammar is called outer syntax.
We developed a parser for the outer syntax of the Isar language, such that user input is immediately checked for syntax errors without being sent to the backend for verification.
The parser is also used for syntax highlighting and restriction for specific courses.

The web client communicates with the web server via HTTP requests and WebSockets for user authorization and sends user input to the server for semantic verification.
The primary functions of the \emph{web server} are: handling authentication and authorization, managing and distributing data, and facilitating communication with the ISM.
For security and privacy reasons, \ProofBuddy uses role-based access control such that, depending on their role, users are authorized to access the intended resources and functions.
The identity and access management is realized with Keycloak\footnote{\url{https://www.keycloak.org}}.
Therefore, \ProofBuddy does not need to store any sensitive data, only the username, the openid issuer (the Uniform Resource Identifier (URI) of the Keycloak instance) if the user has admin rights, and the creation date.
Keycloak also facilitates the integration of single sign-on services, and thus supports the possibility for students to log into \ProofBuddy with their university login credentials.
The user profiles, as well as the materials, including the user progress, are managed in a PostgreSQL\footnote{\url{https://www.postgresql.org}} \textit{database}.

To provide the opportunity to analyze progress and possible mistakes of students, a diff is stored in a database table for each user submission, containing the changes made in comparison to the previous user submission, along with its timestamp.
The diffs are linked to the corresponding user by their user ID.
The reference is needed to map future submissions to the respective user and to evaluate a series of changes from the same user.
After deleting a user profile, the diffs of that user can still be grouped based on the user ID, but since the profile no longer exists, there is no longer a link to any user data, which effectively anonymizes all collected data.

An important part of \ProofBuddy is the \textit{Isabelle Server Manager (ISM)}, a (basic) load balancer and a supervisor for managing Isabelle server \cite{wenzel2014isabelle} instances.
\ProofBuddy integrates (multiple) Isabelle servers in the backend, providing online access to the Isabelle prover without the need of installation by users. The backend integration, as opposed to having it run locally within the browser, is necessary because Isabelle, written in ML and Scala, cannot be easily converted to JavaScript.
Furthermore, external tools used by Isabelle, such as automated provers for smaller problems, prevent full operation within a web browser.
Within \ProofBuddy, users will be unable to utilize automated provers in order to get a better understanding of the individual proof steps.
However, automated provers can be useful for teachers to quickly check whether a task is provable.
To check the correctness and soundness of theories, Isabelle provides the Isabelle server, an integral component of the Isabelle distribution.
The server functions as a daemon and is able to manage multiple Isabelle sessions, a primary concept within Isabelle.
Sessions comprise collections of Isabelle theories and their associated states.
The evaluation of theories occurs within sessions, analogous to projects in IDEs.
Sessions are created in a hierarchical structure, based on parent sessions, with \emph{Pure} \cite{wenzel2014isabelle} being the root of all sessions.
The server architecture permits remote access to an Isabelle instance.
The Isabelle server is engineered to handle concurrent requests, facilitating tasks such as initiating new sessions or adding new theories to existing sessions.
The ISM is responsible for managing a scalable pool of isolated and bounded Isabelle servers.
At startup, it initializes multiple Isabelle server instances, two for courses with up to 25 students, each monitored for metrics such as the number of active sessions.
% scalability
So far, we have successfully tested ISM with up to 30 servers.
While it is important to have enough Isabelle servers to handle the workload efficiently, it's equally crucial not to use more Isabelle servers than required:
Each additional server adds a memory overhead, which can impact the system performance in stress test scenarios.

\subsection{Teaching with \ProofBuddy}
\label{sec:teach-with-proofb}

As in Subsection~\ref{sec:teaching-with-proof}, the arguments of using software as teaching support also applies here.
With \ProofBuddy, learners can start directly without having to install Isabelle themselves.
\ProofBuddy has a functionality similar to Isabelle/jEdit, with an interface for beginners that has the ability hide some syntax constructs (the self written parser is in the frontend) and windows which are only important for experts.
\ProofBuddy's UI design allows students to focus on the proofs and feedback, including the ProofState, from Isabelle.
Since Isabelle is not integrated into \ProofBuddy's frontend a request has to be sent to the server.
We decide to let user actively ask for feedback to decrease trial and error attempts.

%\subsection{Usability Study}
%\label{sec:exper-2-eval}

\begin{figure}
\centering
\includegraphics[scale=0.2]{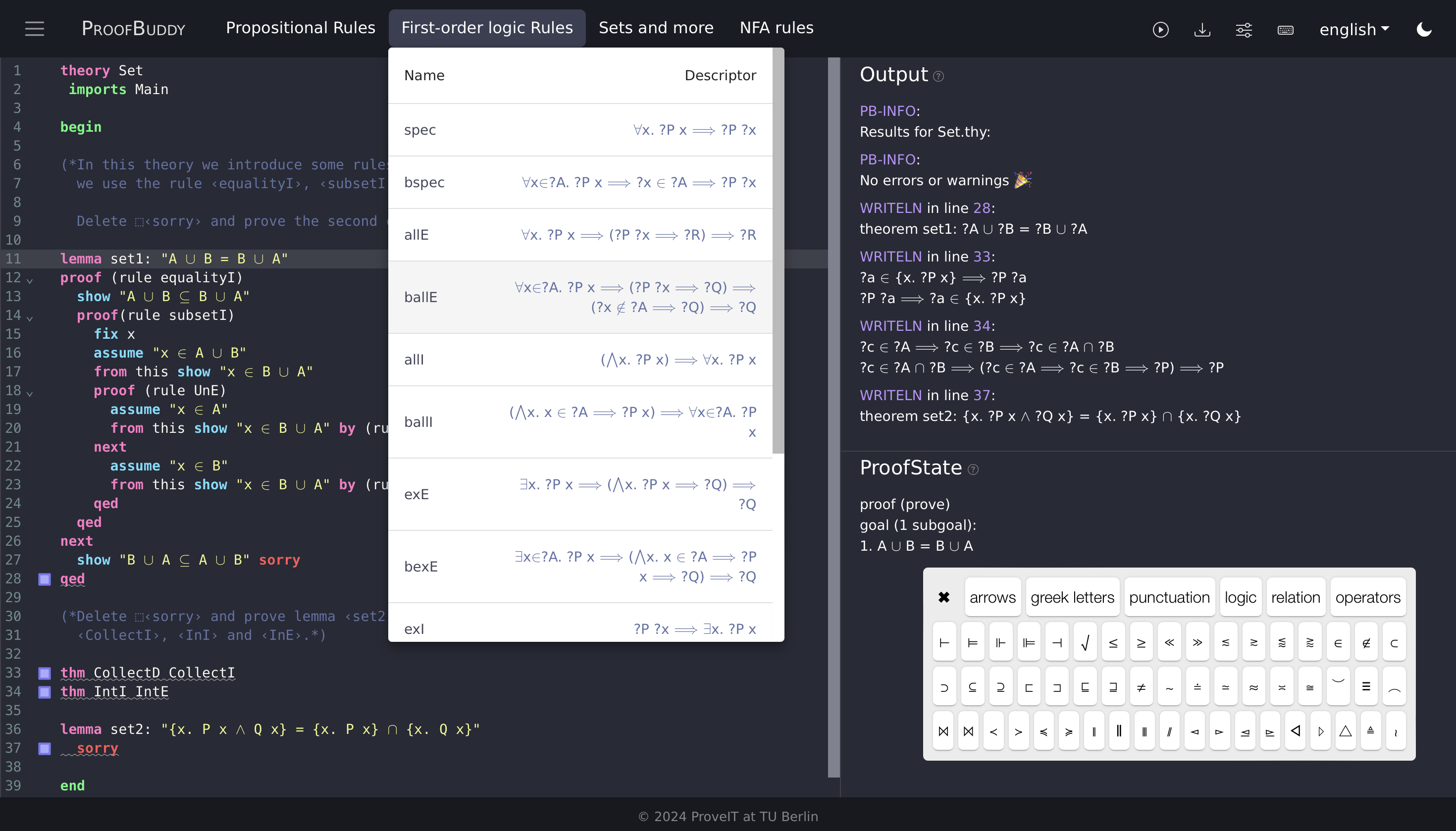}
\caption{Screenshot of \ProofBuddyv{0} (Usability Study)}
\label{fig:pb-v0}
\end{figure}

In spring 2023, we conducted a usability study \cite{proofbuddy2023} employing \ProofBuddyv{0} (\Cref{fig:pb-v0}) within a single session of a course dedicated to learning Isabelle at the Technical University of Denmark (DTU).
The participants in the study were Master's students in CS, all of whom were already familiar with Isabelle/jEdit.
They were tasked to perform an Isabelle exercise using \ProofBuddyv{0}.
The aim of the study was to investigate their interaction with the interface and identify aspects that they found beneficial or detrimental.

In \cite{proofbuddy2023}, we found out that students like to interact with \ProofBuddyv{0}.
They like the interface and especially the option to look up rules such that they did not have to remember each relevant rule (\Cref{fig:pb-v0}).
This figure also shows an on-screen keyboard for Isabelle symbols.
Due to some server problems not all students used \ProofBuddyv{0} for the whole session.
As the complexity of the syntax was not hidden in \ProofBuddyv{0}, students still have a steep learning curve to learn the relevant part of the Isar language.
In addition, Isabelle's feedback is difficult for beginners to interpret, often leaving them feeling overwhelmed.
This study indicates that, while \ProofBuddyv{0} apparently had some advantages over Isabelle/jEdit for our educational context, it was still not yet good enough.

\section{Educational Design I}\label{sec:eduDeI}

We develop our use of Isabelle and later \ProofBuddy via an iteration of a test course for TLPP to continuously improve both the integration \ProofBuddy and the teaching materials, called EDR (\Cref{sec:introduction}), which we will pursue in \Cref{sec:eduDeI,sec:eduDeII}.
Therefore we use a completely new course tailored to test end and evaluate it, which we offer irregularly.

\subsection{Experiment}\label{sec:exp1}

For the summer term 2023, we designed a Bachelor course aimed at CS students from the second semester onwards.
A total of 14 CS students started and all completed the course.
One half of the participants was in semester 2, the other half in semester 4 and above.
In \cite{proofbuddy2023} we plan to use \ProofBuddyv{0} for the first iteration of WoP.
At the end we chose Isabelle/jEdit \cite{IsabellejEdit}  for two reasons: (1) the infrastructure of \ProofBuddy, like a sufficiently scaled server for the number of users, was not completed on time and (2) using Isabelle/jEdit for the first iteration enables us to compare it with \ProofBuddy in the next iteration of the course with nearly the same course materials.
The goal was to have students---with the help of Isabelle---learn \emph{how to structure and write proofs} in natural language.
More precisely, our intention was \emph{not} to have the students just learn how to develop proofs \emph{within a PA}, but rather to use the assistance and feedback of a PA to learn \emph{how to structure and write proofs} such that they could afterwards also do it properly on paper and using natural language instead of the PA language.
We named it ``World of Proofcraft'' (WoP).
The design of the course required two important decisions: (1)~Which topics should be covered in a first year course? and (2)~How to prepare the topics in Isabelle?
The next paragraphs discuss our answers to these questions.

\begin{figure}
\centering
\includegraphics[scale=0.18]{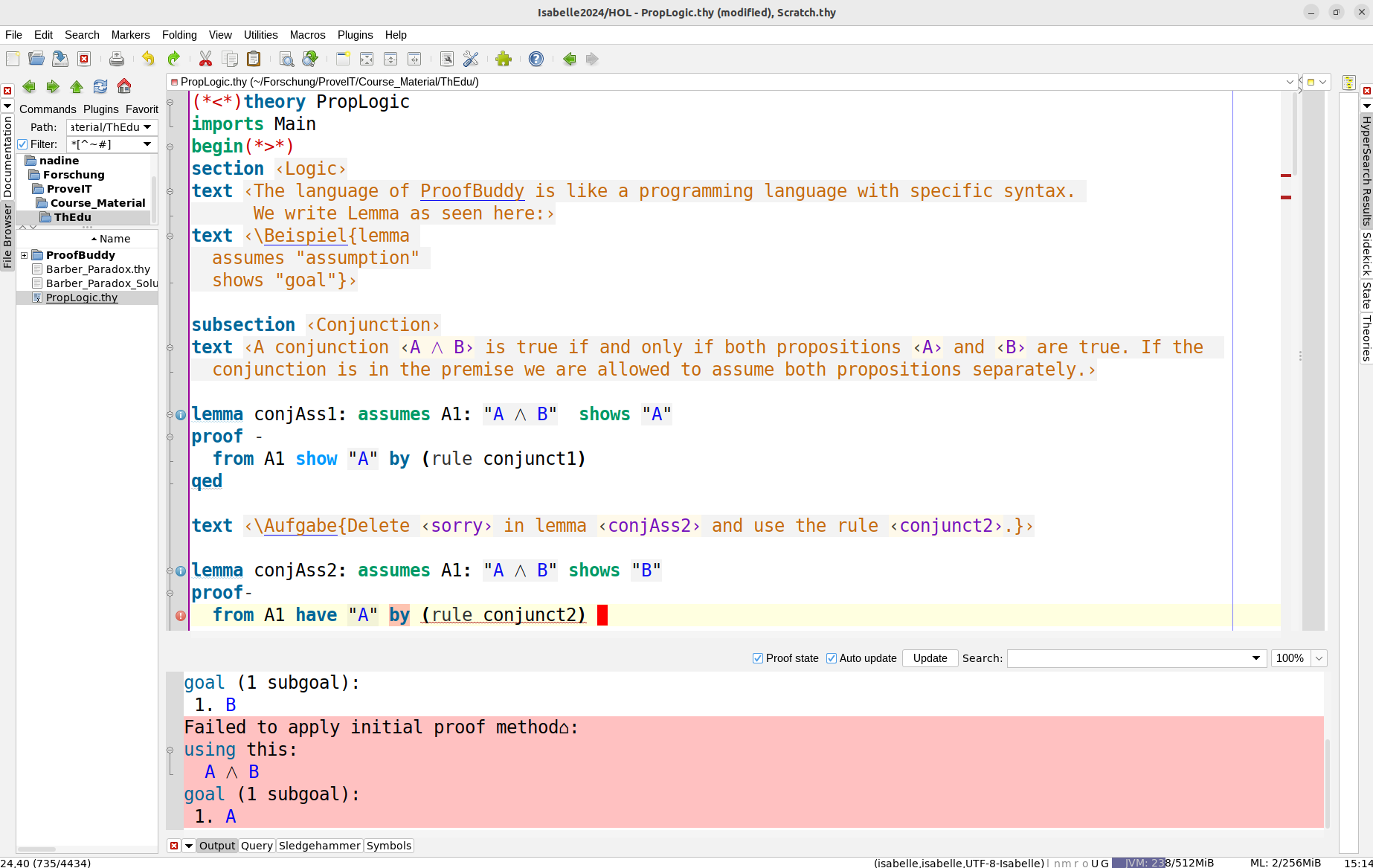}
\caption{Screenshot of Isabelle/jEdit \cite{IsabellejEdit}} (WoP: Iteration 1)\label{fig:jEdit}
\end{figure}

The contents were delivered in 14 weeks with 4 hour lessons per week, using a mixture of classroom teaching and exercise sessions.
WoP covers the following topics: introduction, propositional and first-order logic, proof techniques, types and lists, student projects and presentations.
Each logical connective is introduced with varying degrees of formality \cite{Boehne2019}, similar to the approach by Avigad~\cite{Avigad2019}.
Since the main goal of WoP is to improve proof writing in natural language, we focus on this.
To build a bridge between natural and PA language we use our logical language---natural deduction.
It also helps to prepare students to get acquainted with pattern matching: deduction rules are schematic in that meta variables are used for formulae in premises and conclusion that students need to match with proof states.
A similar pattern matching is also needed to match the rules against the syntax of the Isar language.
The integration of Isabelle should serve as a tool for the students to check their proofs and to get more and instant feedback.
Exercises are conducted in natural, logical and the Isar language and the transfer between these languages is dealt with explicitly.

To gain a better understanding of the struggles that students face when they use a PA and to explore if the described teaching method is helpful for learning how to structure and write proofs, WoP also features so-called \emph{learning portfolios}.
Every week, students have to write a personal report about the content of the course, to document their struggles and to describe the strategies they employed to address them.
Subsequently, students engage in peer review: offering feedback on two other reports and receiving feedback from two peers.
At the end of the semester, they author a reflection document regarding their own overall learning process.
This concept is inherently interconnected, with reflection serving as the main point, encompassing both proof construction and individual learning experience.

%\pagebreak
\subsection{Evaluation}
\label{sec:evaluation-1}
Following the EDR approach, we evaluated our experiment to improve the choice of PA, course topics and material.
In our case, we had to be aware that we are researcher and teacher in one person.
The limited cohort of participants enabled close supervision and a qualitative evaluation of teaching effects.
Initially, we conducted an associated observational study, in which we documented the questions that students directly asked the teacher during course sessions, akin to \cite{KnobKreitz2017}.
Additionally, the questions and reflections of learning portfolios enabled a qualitative evaluation of challenges encountered using qualitative contextual analysis~\cite{Mayring2014}.
%, with the assistance of MaxQDA software for summarization purposes.
To measure improvements in proof structure, errors from pre- and post-tests where students had to write proofs in natural language, were scrutinized.
We combined these analyses using a mixed-methods approach \cite{BaurMethoden} to provide a comprehensive assessment.

To analyze the questions during the course sessions and the problems described in the reflections, we formulated a system of categories inspired by \cite{KnobKreitz2017} to capture the various challenges students encountered, as expressed in their requests for support during the sessions.
The identified categories are the following: creating proofs, writing proofs, mathematical language, logic, Isabelle usability and transfer between natural, logical and Isar language.

\begin{figure}
\centering
\includegraphics[scale=0.055]{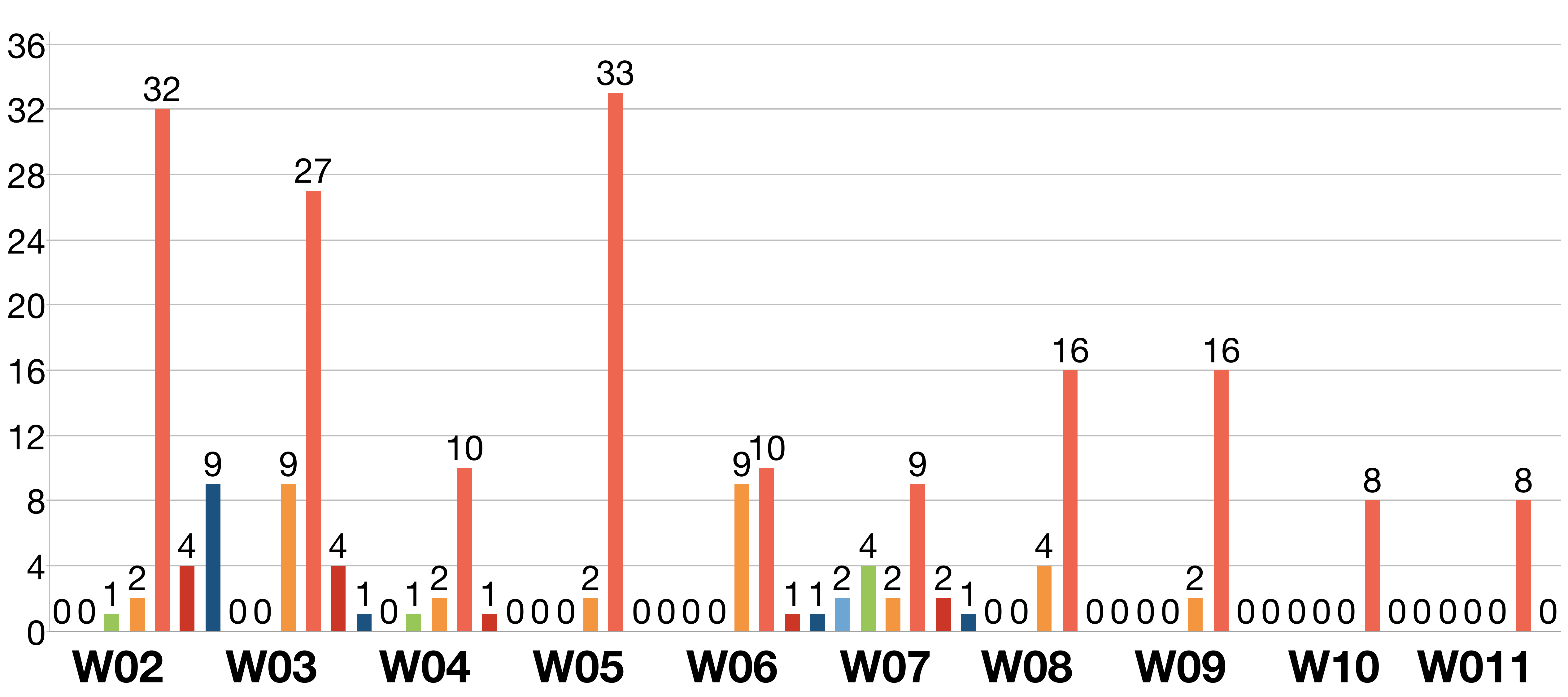}
\caption{Number of the Students' Questions during the Seminar Sessions (WoP: Iteration 1) \label{fig:questionsWoP1}}
\end{figure}

\begin{figure}
\centering
\includegraphics[scale=0.055]{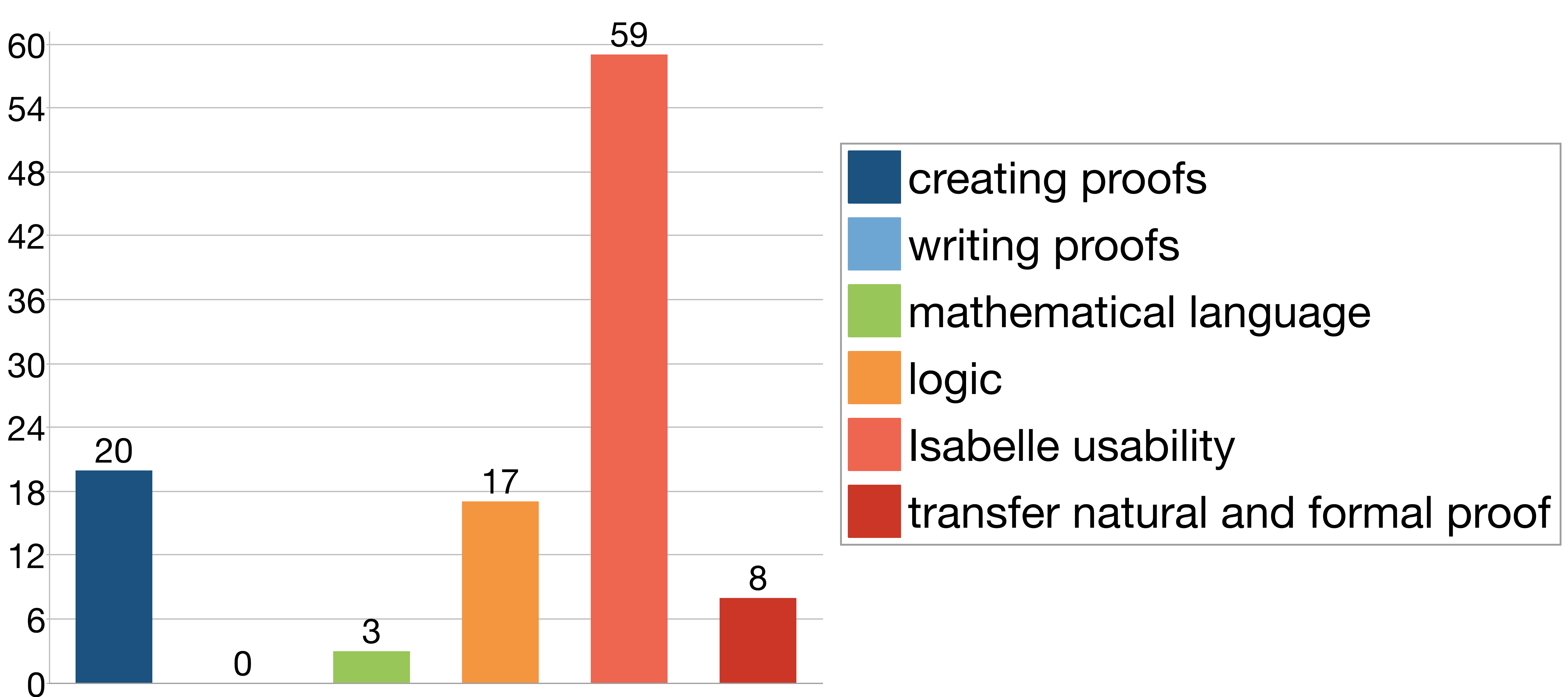}
\caption{Number of described Problems in the Reflection (WoP: Iteration 1) \label{fig:reflectionsWoP1}}
\end{figure}

The number of questions and problems encountered in the course sessions and reflections are represented in \Cref{fig:questionsWoP1} and \Cref{fig:reflectionsWoP1}.
The questions depicted in \Cref{fig:questionsWoP1} are chronologically organized by week, where the legend is the same as in \Cref{fig:reflectionsWoP1}.
Both figures show that there is a predominant focus on Isabelle's usability.
This trend aligns with the course's progression, where each week introduced additional syntax and semantics of the proof assistant, escalating the complexity faced by students.
A significant challenge identified was the matching of assumptions and goals to rules in Isabelle, a difficulty also highlighted in the weekly Isabelle reports, prompting a dedicated session to address this concept specifically.
Consequently, in week 05, we focused on this difficult matching, which caused many questions asked about Isabelle.
This repetition and clarification was also mentioned positively in the reflections.
Furthermore, students had difficulties in reading the error messages of Isabelle which originally were made for expert users and are mostly not sufficiently informative.
Unfortunately, many students skipped the attempt to write a proof in natural language first and directly developed within Isabelle.
Therefore, the number of questions belonging the category ``creating proofs'' decreased over the weeks.
Yet, at the end of the course, most students reflect on their ability to write natural language proofs (\Cref{fig:questionsWoP1}).
During the course, we discussed the use of logic and its connection to proof structures.
In their reflections, most students recognize that the understanding of the logical connectives was important for the understanding of proof structures; they also expect that this would help them later on in other courses, where they would have to write proofs in natural language.

The evaluation shows that the students have a lot of fun programming their proofs and are able to understand the difficulties of structuring proofs.
But they struggle with the complexity of Isabelle especially with the rule matching, the large amount of syntax and the error messages.
The experiment indicates that a PA can help students in learning how to prove because of the immediate feedback.
Improving our course material to focus on the matching of the assumptions and goals of Isabelle's rules is a challenge.
To decrease the initial difficulties, we need to reduce the complexity of Isabelle's syntax, leading us to use \ProofBuddy for the next WoP iteration.
This web application, a now better developed software that shades the power of Isabelle, ensures the possibility to share exercises with student, react on bugs fast and collect data.

%%% Local Variables:
%%% mode: latex
%%% TeX-master: "main"
%%% End:

%% file: interactive-tutorials.tex
\section{Interactive Tutorials}
\label{sec:inter-tutor}

The usability study in Subsection~\ref{sec:teach-with-proofb} indicates that the learning curve for the use of \ProofBuddyv{0} is still somewhat too steep, although less so compared to the typical PA off-the-shelf, we want to use so-called tutorials to better guide students along a didactic learning path.
There are many definitions and interpretations of the word ``tutorial''.
%
%\UN{The following two sentences could also be skipped.}\NK{I would delete them}
%For example, according to the Webster’s New World College Dictionary~\UN{reference?}, a tutorial is ``an intensive course given by a tutor or professor for one or several students, usually on a special topic''.
%
%Or, according to the Collins COBUILD Advanced Learner’s Dictionary~\UN{reference?}, a tutorial can be understood as ``part of a book or a computer program which helps you learn something step-by-step without a teacher''.
Here, we refer to \emph{tutorials} as ``instructional materials designed to help users understand and implement concepts, providing step-by-step instructions, explanations, and examples''.
Each tutorial should have a clear learning goal and structured content.
The aim is to have learners acquire competence for a particular skill set in a more or less specialized area, in a brief and concise manner.
In particular, tutorials are useful to introduce learners to a new topic, for which tailored explanations can be offered.

Providing \emph{interactive} elements in the tutorials allows users to actively engage with the content and enhance their learning experience by reinforcing understanding through ``hands-on'' exercises.
So, learners are confronted with tasks that they shall try to resolve on their own.
The level of interactivity then varies according to the number and granularity of tasks.
Moreover, the learning experience varies greatly with respect to the quality of the user interface (UI), especially concerning the presentation of read-only text (for introductions and explanations) in alternation with writeable tasks.
%, and the support for learning the syntax of the PA language.\NK{the last sentence is the first where PAs are used, for me there is a gap}

\subsection{Teaching with Interactive Tutorials}
\label{sec:teach-with-inter}

Tutorials can be realized using both desktop and web applications.
There are surprisingly few research articles, e.g.~\cite{pop02:comparwebapplicdesktapplic}, (instead of mere blog entries) on comparing these two types of application.
Moreever, there are specific requirements and desirable features when teaching in tutorial style.
Thus, in this subsection, we summarize our own comparative analysis.

The advantages of web applications over their desktop counterparts by avoiding the local installation overhead for learners have already been highlighted above.
In the context of tutorials, these advantages becomes even more striking:
teachers must be really sure that all learners use the same version of the software and they also want to correct and update the software much more frequently and on shorter notice.
This may be to quickly improve either the interface or even the contents of a tutorial (see below).
It is much easier to do so with only a single point of change instead of having all learners update their local installations.

Some requirements for educational software to support tutorials can be fulfilled by both desktop and web applications.
For example, learners typically want to able to save their state of progress to pause and continue later on.
With a desktop application, this is usually trivial.
Web applications would either need to offer the registration of learners on the server and provide storage capabilities there.
Or, they would need to support the download of progress states to local file systems
% (to possibly be able to continue locally with a desktop version, once learners reach a sufficient mastery level)
and to upload such files again.
% that learners have prepared locally (possibly beyond the mandatory part of the tutorial at hand).
Specific features such as, for example, tracking the progress of a learner to display his/her individual achievements are conceived to incentivize learners to complete tasks and to make tutorials more attractive.
Such gamification is also possible with a desktop application, but easier to achieve with a web application.

Clear advantages of web over desktop applications are the following:
As indicated above, the instant and adaptable distribution and availability of a tutorial (or teaching material, in general) is a big plus.
In addition, the software itself may be adapted specifically to the needs of individual tutorials by, for example, changing the interface to shade some functionality for didactical purposes.
With web applications, it becomes much easier to track the activities of learners, which itself enables various actions:
Learners may, in addition to the above-mentioned gamification, get more precise feedback.
Teachers may profit from ``crowd feedback'' by quickly spotting problems that are shared across individual learning paths within the same tutorial.
Developers may be informed about potential interface misconceptions by learners.

\subsection{Interactive Tutorials with Proof Assistants}
\label{sec:inter-tutor-with}

In this section, we discuss tools related to proof construction, for which (interactive) tutorials have been developed.
As we have argued in Subsection~\ref{sec:teach-with-inter}, web applications are superior for this purpose, so we focus on them.
In addition to the rather abstract remarks on general interactive tutorials of the last section, we first mention some concrete possibilities in the context of proof construction.
For example, a web interface for a PA makes it possible to constrain access to full PA language, as a tutorial-specific parser can easily be integrated within the front end.
Likewise, the access to automated tactics of a PA can be prohibited for parts in which learners are to prove obligations explicitly.
A web interface can also, according to the stated learning objectives of a tutorial, control the usage of automatic theorem provers that may normally be accessible from within a PA (as, for example, is the case with Isabelle).

For the proof assistants Coq and Lean, web interfaces have been developed to replicate functionalities akin to their desktop counterparts.
The following paragraphs provide an overview.

Coq has been fully translated into JavaScript with the creation of \emph{jsCoq}\footnote{\url{https://jscoq.github.io/scratchpad.html}} \cite{GallegoArias2017}, allowing direct access through web browsers.
Although not explicitly designed for teaching proof construction, jsCoq excels in presenting mixed-content documents combining readable text with executable Coq code.
However, while users can input (not upload, just copy-and-paste) their own Coq files, the application lacks file management capabilities for storing files in the backend.

There is a \emph{Lean tutorial} interface\footnote{\url{https://leanprover.github.io/tutorial/}} that provides an introduction to the proof assistant.
The interface combines readable text with exercises that can be selected and then appear in the editor window.
Uploading files is not possible in this application.

Another \emph{Lean tutorial} interface is the Lean Game Server\footnote{\url{https://adam.math.hhu.de/}}.\@ This website offers several interactive tutorials for Lean, including the Natural Number Game\footnote{\url{https://adam.math.hhu.de/\#/g/leanprover-community/nng4}}, with which users can learn Lean step by step.
It is possible to create its own game or tutorial and add it to the server.
There is a github repository where a template and an explanation can be found to create a fixed structured web game.
Students can use this platform for free and submit their data to the Lean Game Server, but it is not possible to collect data as a course designer.

\emph{Edukera}\footnote{\url{https://edukera.com/}} \cite{edukera2016}, a commercial web application, integrates Coq for proof verification.
Although intuitive, it can take a considerable amount of time for users to familiarize themselves with it, as mentioned in \cite{wemmenhove2023waterproof}.
The interface is organized into courses, providing explanations in readable text along with exercises, resembling a tutorial format.
However, users are not actively \emph{writing} proofs themselves, but rather click on possibilities or suggestions.
Thus, this tool may not align with the intention of having students learn how to use any \emph{language} to write proofs themselves.
Additionally, the absence of an option to upload custom files further limits its utility.

Most of the above-mentioned learning interfaces lack the ability to upload custom materials, and none of them officially offer data collection features, so they can not be used to analyze user interactions to improve the learning material.
In contrast, all of this is made possible by the \ProofBuddy web application, as we will see in the next section.

\subsection{Interactive Tutorials for \ProofBuddy}

\begin{figure}
  \centering
  \includegraphics[width=0.9\linewidth]{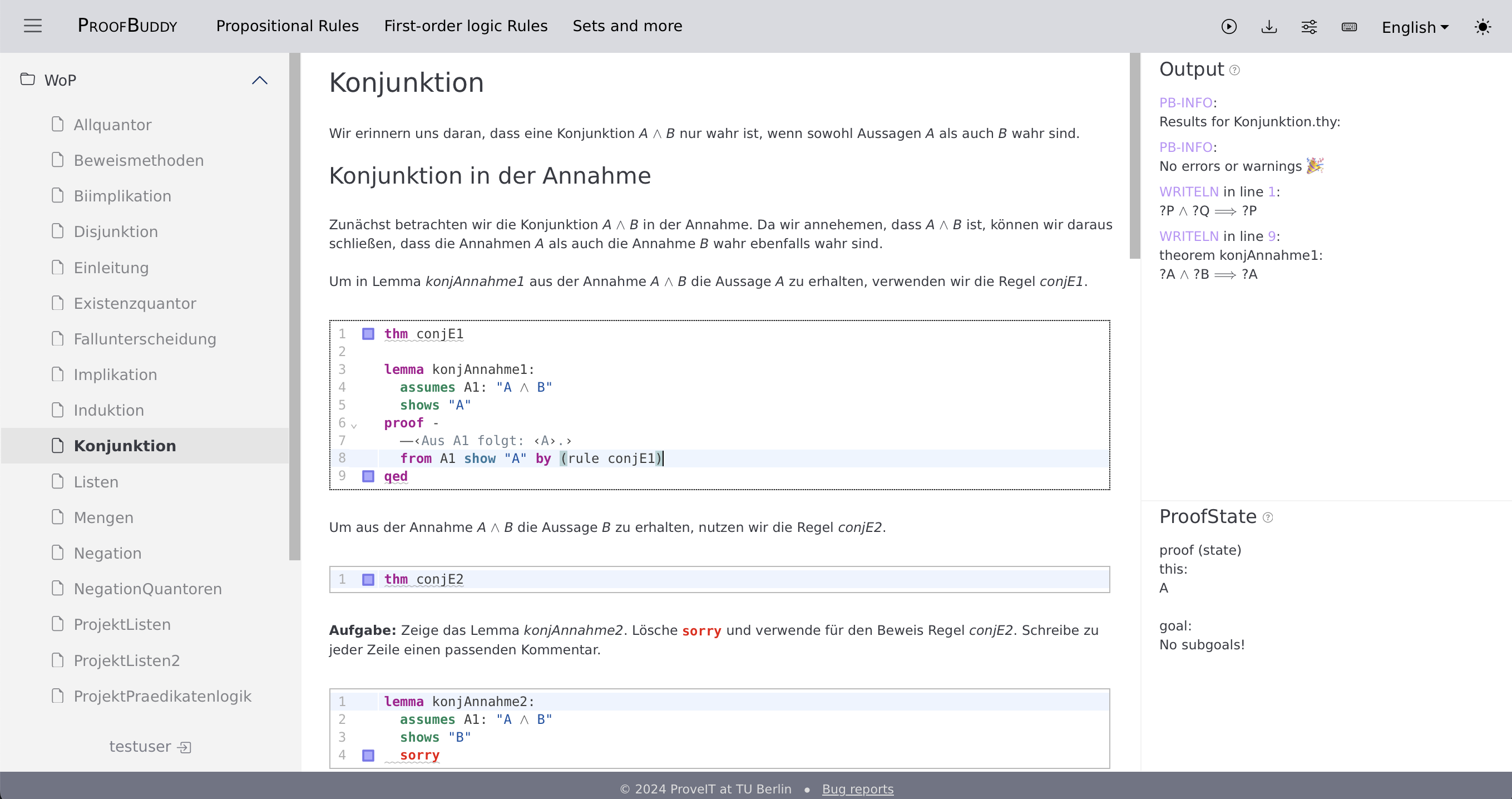}
  \caption{Screenshot of an Interactive Tutorial in \ProofBuddyv{1} (WoP: Iteration 2)}
  \label{fig:pb-v1}
\end{figure}

\begin{figure}
  \centering
  \includegraphics[width=0.9\linewidth]{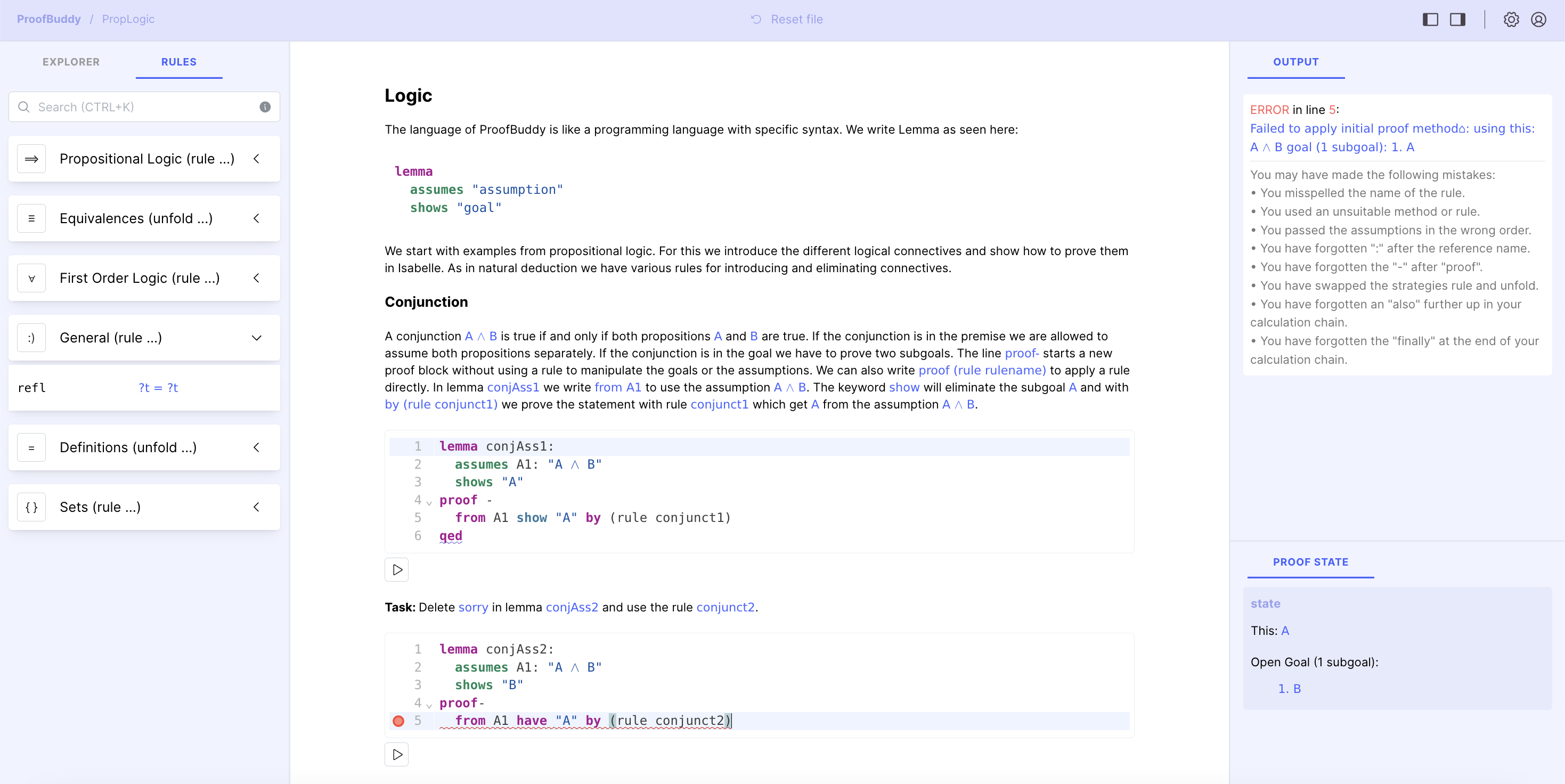}
  \caption{Screenshot of an Interactive Tutorial in \ProofBuddyv{2} (WoP: Iteration 3)}
  \label{fig:pb-v2}
\end{figure}

\ProofBuddy provides interactive tutorials (see, for example, \ProofBuddyv{1} in \Cref{fig:pb-v1}, or \ProofBuddyv{2} in \Cref{fig:pb-v2}) to facilitate the TLPP.
In \ProofBuddy, users can find editable and executable tutorials that are grouped into courses.
% tutorials
The tutorials combine textual elements, examples and tasks to guide students through the process of constructing proofs, similar to the tutorials in \emph{Waterproof}\footnote{\url{https://github.com/impermeable/waterproof/blob/develop/package.json}} \cite{wemmenhove2023waterproof}, which is a VSCode extension based on Coq and enables a presentation style resembling paper proofs.
In the tutorials of \ProofBuddy, students are asked to write Isabelle code to, for example, define or prove given statements.
The Isabelle code can be written in several small editor components, similar to jsCoq\footnote{\url{https://jscoq.github.io/scratchpad.html}} \cite{GallegoArias2017}, Waterproof and Jupyter Notebook\footnote{\url{https://jupyter.org/}}, instead of having students work on multiple exercises within a continuous editor, as in Isabelle/jEdit.
Each editor instance is dedicated to a specific \textit{task}.
%This allows students to focus on one task at a time and prevents them from being overwhelmed by the complexity of the entire theory.
This allows for a more structured and focused learning experience, as students can concentrate on one task at a time.
The \textit{textual components} allow teachers to explain proof strategies and concepts in a pure text format, rather than being limited to comments within a block of code.
\emph{Examples}, i.e. Isabelle code that is \emph{not} intended to be edited by users, can be displayed in separate read-only editor instances.
The read-only editors can also be used to display definitions and lemmas provided by teachers as a working basis for students.
In the background, each tutorial represents a single Isabelle theory divided into smaller parts that will appear to the user as separate proof tasks.
Therefore, the tasks within the tutorials can be presented in a step-by-step manner, with the content of each task building on the previous one.
Additionally, mandatory definitions are integrated into the theory in the backend, such that students are unable to alter them.
This also means that those parts of Isabelle theories that are not needed for the formulation of proofs, but are essential for Isabelle theories, can be hidden from the user.
Examples include the theory context or the imports of other theories.
For convenience, users are able to reset their progress in a tutorial, allowing them to start from the beginning if they wish.

% feedback
At the user's request, the Isabelle prover is used to generate feedback on the correctness of the tasks, as follows:
With the current user input, a theory is reconstructed from the tutorial, including the hidden and mandatory sections, and sent to the Isabelle server for verification.
The Isabelle server provides feedback on errors, warnings and the proof state.
However, since the feedback from Isabelle can be generic and difficult to understand for beginners, \ProofBuddy provides additional hints on possible mistakes that caused the error.
As an example, \Cref{fig:pb-v2} shows how possible mistakes that lead to the failed application of a proof method are outlined.

Errors and warnings are displayed to the user in the \textit{Output tab} of \ProofBuddy.
In addition, the corresponding sections of code are highlighted in the editors, with the feedback messages displayed when hovering over them.
Isabelle's feedback is clearly labeled to ensure credibility and transparency.

In the \textit{ProofState tab}, the user can view the ProofState to gain a better understanding of the constructed proof. The ProofState, including the subgoals still to be proven, is displayed for the active line in the currently selected editor instance.
Syntactic feedback is provided instantaneously by the custom parser that is integrated within the frontend of \ProofBuddy.
The parser also allows for syntax restrictions and prohibition of certain rules and keywords, e.g., for the automatic Isabelle tactics such as \texttt{auto}, \texttt{simp} and \texttt{blast}.
Linters then display the restrictions or issue warnings in the editor, reminding users to better choose alternative language constructs or tactics.

% ui components
In order to let users focus on the construction of proofs, \ProofBuddy offers a \textit{Rule tab} listing all available Isabelle rules relevant for the specific teaching context as well as a search option.
This allows users to browse and select rules directly for their proofs.
Additionally, the \textit{symbol lookup} contains a search bar and list of all usable symbols with their name and abbreviations (e.g. the symbol $\wedge$, named `and', with abbreviation $/\backslash$).
An auto-complete function suggests possible completions for the current input, including symbols and their abbreviations, keywords and rules.
Thus, users can write proofs more efficiently and with fewer spelling mistakes.

The \textit{Explorer tab} provides an overview of the courses and tutorials currently available, from which users can select any to work on.
The different tabs (Output, ProofState, Explorer and Rules) can be positioned via drag-and-drop individually at the left or right side of the screen, allowing users to customize the interface according to their preferences. For more space, the tab sections on the left and right can be minimized.
\ProofBuddy is currently available in English and German, with the option to add additional languages in the future.

\section{Educational Design II}\label{sec:eduDeII}

As stated in \Cref{sec:eduDeI}, we will iteratively develop and evaluate the use of PAs and especially \ProofBuddy.
We use the outcome of the usability study (Subsection~\ref{sec:teach-with-proofb}) and the evaluation of the first iteration of WoP (\Cref{sec:eduDeI}) to develop tutorials for \ProofBuddy (\Cref{fig:pb-v1}).
In this section we describe this iteration, evaluate it and compare it with the first iteration.

\subsection{Experiment}\label{sec:exp3}

In the summer term 2024, we ran the course WoP for the second time (the first iteration is described in Subsection~\ref{sec:exp1}) with the same type of exam, aimed at first-year CS students who are expected to have a basic knowledge of Mathematics.
In this iteration, 22 students started the course, 21 participated until the last session and 15 submitted the reflection at the end.
The content was almost the same as the first iteration, with some different focuses depending on the students' interest and questions.
Instead of Isabelle/jEdit, we now used \ProofBuddyv{1} (\Cref{fig:pb-v1}) as the proof tool with the unfiltered feedback of the Isabelle server.
Otherwise, the course stuck with the two goals of the first iteration:
\begin{inparaenum}[(i)]
\item to teach students how to structure proofs, and
\item to encourage reflection on their learning processes.
\end{inparaenum}

This iteration spanned 14 weeks including an introduction, propositional logic and first-order logic, proof techniques, types and lists, project work and final discussions.
Compared to the first iteration (Subsection~\ref{sec:exp1}), we spent considerably more time on natural deduction and introduced it---being a logical language---as a ``bridge'' between natural language and Isar.
All students should first write their proofs in natural language or logical language to structure their proof ideas.
In a second step, they should translate this proof into the Isar language to make it treatable by \ProofBuddyv{1}.
The students then had to check for themselves whether their original natural language proof was correct and to also modify it ``on paper'' if necessary.
The minimal presentation (hiding some syntax) of the Isabelle files within \ProofBuddyv{1} (\Cref{fig:pb-v1}) let students concentrate on the respective tasks at hand as well as on the given explanations.
According to the inference rules of logical language, we relabeled some rule names of Isabelle in a hidden file to make them easier to identify and closer to the logical language rules.
We also reduced the possibilities of writing proofs in \ProofBuddy~v{1}.
Thereby, students had only one pattern per operator.
This also reduced the complexity of the syntax.

\subsection{Evaluation}

To be consistent with the EDR (\Cref{sec:introduction}), we were also evaluating this iteration of the course to improve \ProofBuddy and the course material.
This time, we especially had to analyze the aspect of focusing more on logical language and on reducing the complexity of Isar language.

For comparison, we chose a similar setting for the study as in Subsection~\ref{sec:evaluation-1}.
We had a pre- and post-test, a survey at the end, and again the learning portfolios.
The collection of questions that students asked during the course sessions was expanded.
We additionally documented where the students pointed at, what the teacher's answer was and what medium was chosen for the explanation.
To do this, we recorded the teacher's voice as well as the student's questions.
In the first two sessions, the students were slightly confused by this (for them) unusual recording activity, but they soon got familiarized with the use of a microphone and the fact that a person who is not involved in the course ``follows'' the teacher.

\begin{figure}
\centering
\includegraphics[scale=0.1]{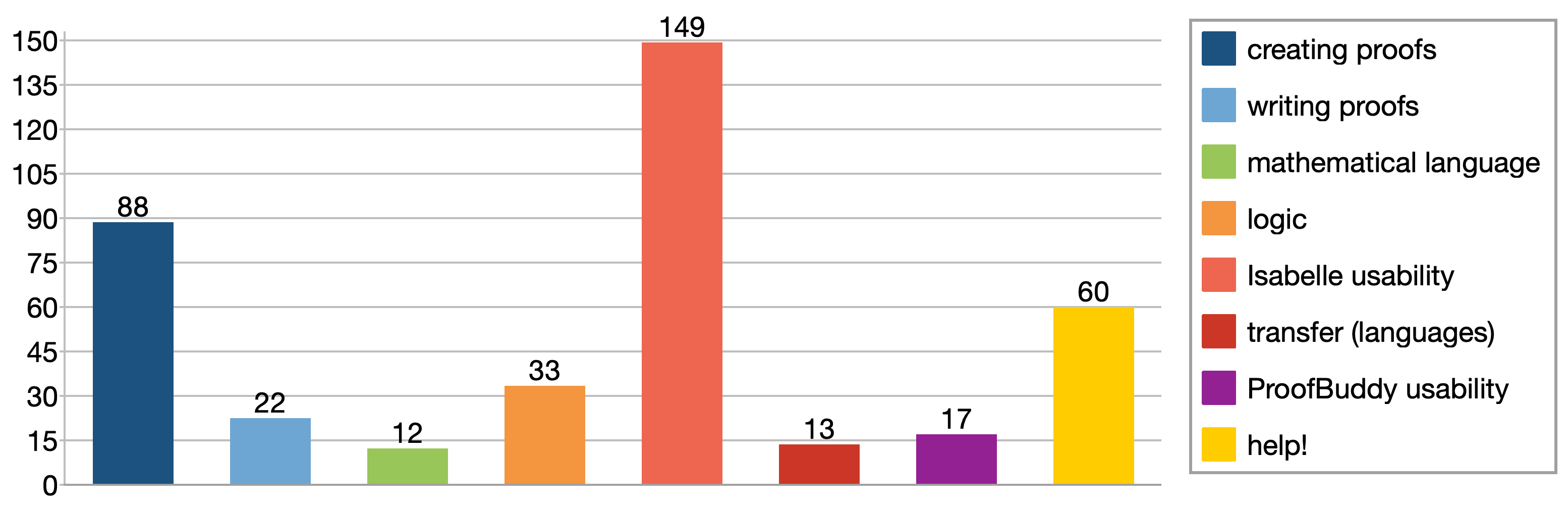}
\caption{Number of Students' Questions During Course Session (WoP: Iteration 2)\label{fig:wop2All}}
\end{figure}

In the following, we evaluate the number of questions that the students asked during the various sessions of the course.
We also explain the categories with respect to the order of their importance, as deducible from the sheer number of questions.
\Cref{fig:wop2All} shows the total number of \emph{all} questions.
They are clustered roughly as in \Cref{fig:questionsWoP1} (Subsection~\ref{sec:evaluation-1}).
For categorization, we added the category ``help!'' for questions where the students were not able to formulate what their question exactly was and just asked for general help.
There were more questions in this iteration, because there were more participants in the course and the results of the pre-test were worse compared to the first iteration (Subsection~\ref{sec:exp1}), which lead us to expect that the students had comparably less prior experience in proving.

\begin{figure}
\centering
\includegraphics[scale=0.1]{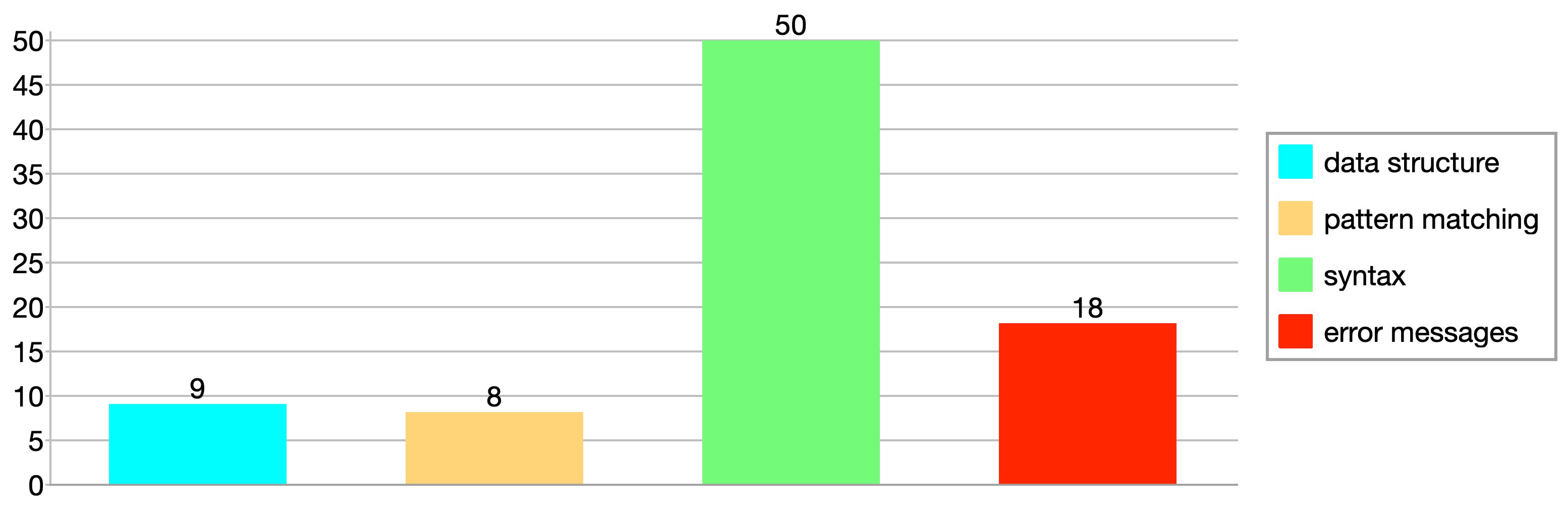}
\caption{Number of Students' Questions during Course Session (Category ``Isabelle usability'' in \Cref{fig:wop2All})\label{fig:wop2Isabelle}}
\end{figure}

\begin{figure}
\centering
\includegraphics[scale=0.1]{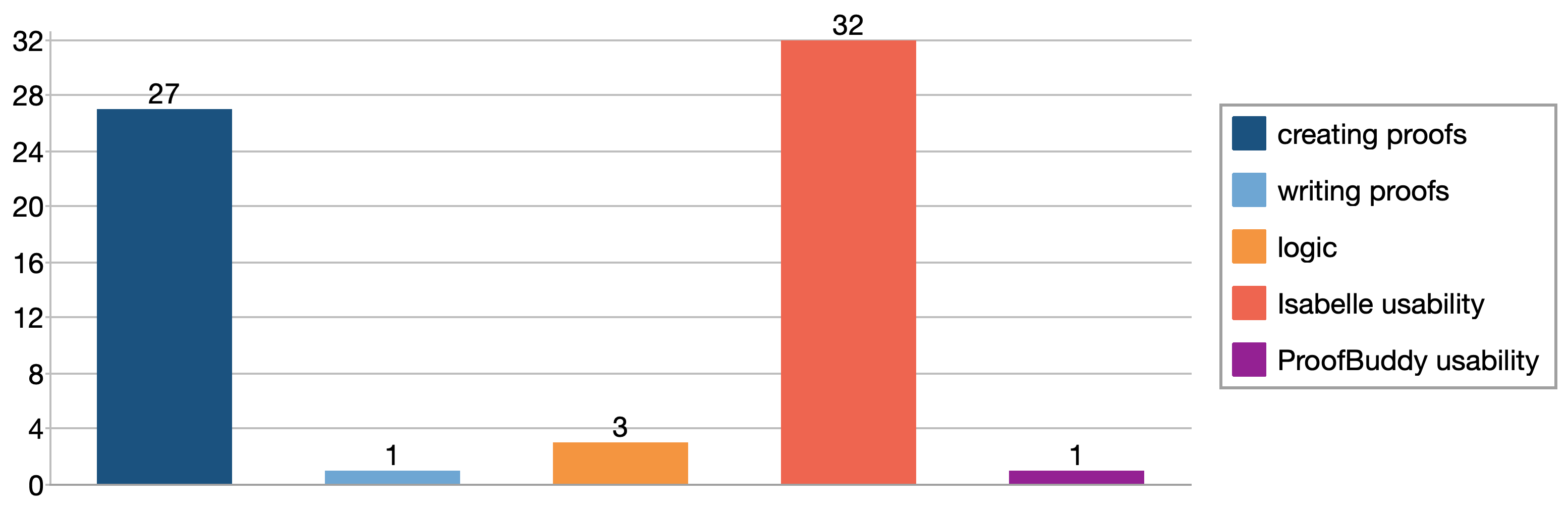}
\caption{Number of Students' Questions During Course Session (Category ``help''! in \Cref{fig:wop2All}) \label{fig:wop2problems}}
\end{figure}

\Cref{fig:wop2All} shows that students still have problems with Isabelle.
We categorized in \Cref{fig:wop2Isabelle} the questions belonging to the Isabelle usability in ``data structure'', ``pattern matching'', ``syntax'' and ``error messages''.
It can be seen that most of the questions fit with the category ``syntax''.
There was one session, where we explained the syntax of forward and backward proofs together with the mapping of rules in Isabelle, many questions about the Isar language and rule matching have arisen.
In \Cref{fig:wop2All} can also be seen that a big difficulty is Isabelle's error messages, which do not help students to understand their mistakes.
In \ProofBuddyv{1}, it is already indicated whether it is an outer syntax (the grammar of the Isar language) error, which belongs to the syntax of the Isar language.
But students need more than just feedback that there \emph{is} an error.
Oftentimes, students get confused as to where the error belongs.

\Cref{fig:wop2All} also shows that students had many questions in the category ``creating proofs'', as they were forced to write down the proofs in natural language or logical language first.
The number of questions in the category ``help!'' is depicted in \Cref{fig:wop2All}, the actual contents is then broken down in \Cref{fig:wop2problems}:
In this category, the usability of Isabelle including the understanding of Isabelle's error messages caused the most problems, as we have already evaluated before.
The second most category where the questions in ``help!'' belongs to is also ``creating proofs''.

At the end of the course, we had a discussion about the course content, the material and the integration of Isabelle.
This interaction at the end of the course suggested that students did enjoy the usage of \ProofBuddyv{1}.
Often, the Rule tab is mentioned positively during the discussion, as students do not have to remember all the rules, which is especially useful at the beginning. % where syntax and rules are new.
The post-test showed that students were a bit better at writing structured proofs after the course than they were before.
In the reflection in the end of semester, students mentioned that they are still unsure about writing paper proofs.

From the points mentioned in the previous paragraphs, we derive the following:
We have to focus more on natural language proofs and the transfer between all three languages---natural language, logical language and the Isar language---in the next iteration.
We will tackle the struggle with the Isar syntax by further reducing it and improving its explanation and exercises.
The new material will be used for the next iteration (\Cref{sec:futureWork}).
For better orientation, the sections given in the Explorer tab will be shortened and all exercises will be organized in separate code blocks.
This reduction in complexity and the addition of more and better explanations will---hopefully---better support the spirit of interactive tutorials.

We also started to collect the diffs between the files the students sent to the server to be verified.
This gives us the opportunity to reconstruct the history of all files.
Since students ask for feedback on syntax errors in the Isar language, this data is difficult to use for analyzing proofs and logical errors and thus for the development of tailored feedback and exercises.

%%% Local Variables:
%%% mode: latex
%%% TeX-master: "main"
%%% End:

%% file: conclusion.tex
\section{Conclusion}
\label{sec:conclusion}

In this paper, we describe our journey using the EDU approach using TLPP-experiments to integrate our \ProofBuddy within the dedicated teaching context WoP.
Here, we see \ProofBuddy play a purely instrumental role: it shall serve just as a supportive means to have students learning how to write proofs by themselves---even without the tool!---also in natural language.
Therefore, the interaction with \ProofBuddy should be as smooth and convenient as possible and not be perceived as an additional hurdle. Proving as such is hard enough.

So far, we found that the steep learning curve can be flattened by reducing the syntax of a PA and providing explanations as the syntax is gradually expanded.
Learners also find it good to have an overview of useful rules, especially at the beginning when the PA syntax is also new to them.
We also conclude in this paper that a learning interface that offers the possibility---from the perspective of teachers---to clearly distinguish between explanations and editable code (with definitions and proofs) may be the best option for learning.
Moreover, we conclude that the immediate feedback indeed helps the students, but that the presentation in Isabelle is often unreadable for learners and overwhelms them.

We decided to develop our test course WoP, its course material and \ProofBuddy \emph{iteratively} using the EDR approach.
In each iteration, we may change the material or adapt \ProofBuddy and observe what is improved and what is not.
By now, we have just completed two iterations of WoP and have collected a lot of data providing evidence.
We can make more or less informed guesses and will analyze the data further.
Interestingly, the EDR approach also guides us to employ better motivated data gathering.
Also for this reason, we are running the course in winter 2024  and will repeat it in winter 2025.
Based on the EDR approach, we also further improved the feedback in \ProofBuddyv{2} for the current iteration of WoP (\Cref{sec:wop3}).

\section{Current and Further Work}
\label{sec:futureWork}

In this section, we describe the current and further iterations of WoP and how we intend to use the findings to transfer our knowledge and infrastructure to a freshly designed compulsory course in the second semester of the Bachelor program CS at TU Berlin: in it, \ProofBuddy will be used at large scale to support hundreds of students acquiring proof competence.

We start to improve Isabelle's feedback in our current experiment and plan to provide more individual feedback for learners to facilitate personalized learning.
To get more feedback for teachers, we plan to integrate learning analytics \cite{ifenthaler2016learning,lukarov2014data} for crowd feedback.
For this purpose, we have gathered and stored some learning data that we can already analyze.
We will use all these features in the previously-mentioned larger compulsory course.

Another point is the development of a learning platform that can be easily used by teachers and students.
Starting with a gentle interface for the generation of course material, preparing sample lectures with exercises and improvements concerning the interface for students.
We also plan a tutorial management system with integrated learning analytics.

\paragraph{Current Experiment.}\label{sec:wop3}

In winter 2024, we started the current iteration of WoP using \ProofBuddyv{2} (\Cref{fig:pb-v2}).
16 students, mostly in their third semester, started it and 14 took part until the last session.
At TU Berlin, the winter term spans 16 weeks, so there is more space to cover the topic of functions and relations (compared to the summer term as in \Cref{sec:eduDeI} and \Cref{sec:eduDeII} with only 14 weeks).
The rest of the content is the same as in the two previous iterations, but with even more focus on the logical language, meaning that the students have to write the proofs first in natural language, then transfer it to logical language and after this intermediate step translate the proof to \ProofBuddyv{2}.
The sections in \ProofBuddyv{2} are shorter than in the previous iteration.
More rules are renamed to fit with the names of logical language and rules are presented as in \Cref{fig:pb-v2}:
Here, one can also see that Isabelle's feedback is provided with additional hints as to \emph{where} the error lies.
Due to the EDR approach, we again collect the students' questions during the course and the teacher's answers.
In addition, we store the diffs of the files with timestamps that students send to the Isabelle server for verification.
Until now, there is no evaluation of the categories and number of questions asked during the course sessions as well as the reflection, because the course is still running.

\paragraph{Next Iteration.}

For winter 2025, we plan the last iteration of WoP.
For it, we will integrate some new features to \ProofBuddy and the evaluation method according to the EDR approach.
We will further improve the user interface to make it more task-specific and evoke the aspect of gamification.
Depending on the currently selected tutorial, only the exactly required rules will be displayed, some syntax will be allowed or forbidden, and also the feedback and hints will be tailored to this context.
In addition, further improvements to the \ProofBuddy interface are planned, which include 2D-visualizations of proofs and extended explanations for rules.

Following the initial iteration, we will examine the collected data to determine the appropriateness of the problem categories we have established.
Subsequently, our attention will shift towards enhancing the feedback provided by Isabelle, initially targeting the outer syntax before delving into improvements related to the proof content itself.
Combined with the above-mentioned aspects, this mitigates the disadvantages associated with PAs used in educational contexts and, in our opinion, represents a significant step towards making PAs usable in such settings.

%\subsection{Experiment 4}
%\label{sec:futureWoP}
\paragraph{Acid Test.}

% The visualization of the proofs that we test in the last planed iteration of WoP  with guidance for students is planed to integrate into a compulsory course in the second semester with 500 to 600 students.
As the above-mentioned new compulsory course will have to serve 500 to 600 students, we must harden and expand \ProofBuddy's infrastructure and improve the scalability of its server.
The content of the course will be similar to WoP:
The goal of the course is that students can write proofs in natural language on topics covering discrete structures, but also algebra and formal languages.
We will evaluate this course with qualitative methods.

%\subsection{Tutorial Management System}
\paragraph{Tutorial Management System.}

% \KE{ideas:
% \begin{itemize}
%   \item learning management system (LMS) / virtual learning environment (VLE) / learning platform
%   \item learning content management system (LCMS)
%   \item tutorial environment / tutorial platform / tutorial management system
%   \item interactive + one-of-the-above
% \end{itemize}
% }

We plan to extend our tutorial management functionalities with a convenient interface for teachers to create tutorials with customized features like linters for specific language constructs in \ProofBuddy.
It is also planned that teachers will have more options to, e.g., give hints on needed rules or further explanations.
We are also developing more ways to visualize proofs that teachers can choose.
Tutorials will be organized into courses where teachers can invite and manage users and present context-specific tutorials.
To provide even smaller educational units, tutorials will be broken down into sections.
Thereby, teachers can offer multiple tasks and examples per tutorial, without overwhelming their students.

The progress of each user within a course is going to be presented for the individual user.
This includes gamification elements that we will use to motivate students to advance through successive ``levels'', thereby increasing engagement and motivation.
Additionally, a comprehensive overview of student progress enables teachers and teaching assistants to intervene at an early stage if a majority of students is struggling with a particular exercise.

%%% Local Variables:
%%% mode: latex
%%% TeX-master: "main"
%%% End:

%% file: main.bbl
\begin{thebibliography}{10}
\providecommand{\bibitemdeclare}[2]{}
\providecommand{\surnamestart}{}
\providecommand{\surnameend}{}
\providecommand{\urlprefix}{Available at }
\providecommand{\url}[1]{\texttt{#1}}
\providecommand{\href}[2]{\texttt{#2}}
\providecommand{\urlalt}[2]{\href{#1}{#2}}
\providecommand{\doi}[1]{doi:\urlalt{http://dx.doi.org/#1}{#1}}
\providecommand{\bibinfo}[2]{#2}

\bibitemdeclare{incollection}{Avigad2019}
\bibitem{Avigad2019}
\bibinfo{author}{Jeremy \surnamestart Avigad\surnameend}
  (\bibinfo{year}{2019}): \emph{\bibinfo{title}{{Learning Logic and Proof with
  an Interactive Theorem Prover}}}.
\newblock In \bibinfo{editor}{Gila \surnamestart Hanna\surnameend},
  \bibinfo{editor}{David~A. \surnamestart Reid\surnameend} \&
  \bibinfo{editor}{Michael \surnamestart de~Villiers\surnameend}, editors: {\sl
  \bibinfo{booktitle}{Proof Technology in Mathematics Research and Teaching}},
  {\sl \bibinfo{series}{Mathematics Education in the Digital
  Era}}~\bibinfo{volume}{14}, \bibinfo{publisher}{Springer International
  Publishing}, \bibinfo{address}{Cham}, pp. \bibinfo{pages}{277--290},
  \doi{10.1007/978-3-030-28483-1\_13}.

\bibitemdeclare{book}{BaurMethoden}
\bibitem{BaurMethoden}
\bibinfo{author}{Nina \surnamestart Baur\surnameend} \& \bibinfo{author}{Jörg
  \surnamestart Blasius\surnameend} (\bibinfo{year}{2015}):
  \emph{\bibinfo{title}{{Handbuch Methoden der empirischen Sozialforschung}}}.
\newblock \bibinfo{publisher}{Springer VS}, \doi{10.1007/978-3-531-18939-0\_1}.

\bibitemdeclare{book}{bookCoq}
\bibitem{bookCoq}
\bibinfo{author}{Yves \surnamestart Bertot\surnameend} \&
  \bibinfo{author}{Pierre \surnamestart Castéran\surnameend}
  (\bibinfo{year}{2004}): \emph{\bibinfo{title}{{Interactive theorem proving
  and program development. Coq'Art: The Calculus of inductive constructions.}}}
\newblock \bibinfo{publisher}{Springer Berlin, Heidelberg},
  \doi{10.1007/978-3-662-07964-5}.

\bibitemdeclare{phdthesis}{Boehne2019}
\bibitem{Boehne2019}
\bibinfo{author}{Sebastian \surnamestart B{\"o}hne\surnameend}
  (\bibinfo{year}{2019}): \emph{\bibinfo{title}{{Different Degrees of
  Formality}}}.
\newblock \bibinfo{type}{Phd thesis}, \bibinfo{school}{Universit{\"a}t
  Potsdam}, \doi{10.25932/publishup-42379}.

\bibitemdeclare{inproceedings}{inp:BoehneKnobelsdorfKreitz16a}
\bibitem{inp:BoehneKnobelsdorfKreitz16a}
\bibinfo{author}{Sebastian \surnamestart B{\"o}hne\surnameend},
  \bibinfo{author}{Maria \surnamestart Knobelsdorf\surnameend} \&
  \bibinfo{author}{Christoph \surnamestart Kreitz\surnameend}
  (\bibinfo{year}{2016}): \emph{\bibinfo{title}{{Mathematisches Argumentieren
  und Beweisen mit dem Theorembeweiser Coq}}}.
\newblock In \bibinfo{editor}{Andreas \surnamestart Schwill\surnameend} \&
  \bibinfo{editor}{Ulrike \surnamestart Lucke\surnameend}, editors: {\sl
  \bibinfo{booktitle}{HDI 2016 -- 7. Fachtagung zur Hochschuldidaktik der
  Informatik}}, {\sl \bibinfo{series}{Commentarii informaticae
  didacticae}}~\bibinfo{volume}{10}, \bibinfo{publisher}{Universit{\"a}tsverlag
  Potsdam}, pp. \bibinfo{pages}{69--80}.
\newblock
  \urlprefix\url{https://publishup.uni-potsdam.de/opus4-ubp/frontdoor/deliver/index/docId/9482/file/cid10\_S69-80.pdf}.

\bibitemdeclare{inproceedings}{coqToText}
\bibitem{coqToText}
\bibinfo{author}{Sebastian \surnamestart B{\"o}hne\surnameend} \&
  \bibinfo{author}{Christoph \surnamestart Kreitz\surnameend}
  (\bibinfo{year}{2018}): \emph{\bibinfo{title}{{Learning how to Prove: From
  the Coq Proof Assistant to Textbook Style}}}.
\newblock In \bibinfo{editor}{Pedro \surnamestart Quaresma\surnameend} \&
  \bibinfo{editor}{Walther \surnamestart Neuper\surnameend}, editors: {\sl
  \bibinfo{booktitle}{Theorem proving components for Educational software}},
  {\sl \bibinfo{series}{Electronic Proceedings in Theoretical Computer
  Science}} \bibinfo{volume}{267}, \bibinfo{publisher}{Open Publishing
  Association}, p. \bibinfo{pages}{1–18}, \doi{10.4204/eptcs.267.1}.

\bibitemdeclare{book}{Brunner2014}
\bibitem{Brunner2014}
\bibinfo{author}{Esther \surnamestart Brunner\surnameend}
  (\bibinfo{year}{2014}): \emph{\bibinfo{title}{{Mathematisches Argumentieren,
  Begründen und Beweisen}}}.
\newblock \bibinfo{publisher}{Springer Spektrum Berlin, Heidelberg},
  \doi{10.1007/978-3-642-41864-8}.

\bibitemdeclare{inproceedings}{Earth2023}
\bibitem{Earth2023}
\bibinfo{author}{Steve \surnamestart Earth\surnameend}, \bibinfo{author}{Jeremy
  \surnamestart Johnson\surnameend} \& \bibinfo{author}{Bruce \surnamestart
  Char\surnameend} (\bibinfo{year}{2023}): \emph{\bibinfo{title}{{Proof Buddy:
  A Tool to Aid Students in Proof Construction}}}.
\newblock In: {\sl \bibinfo{booktitle}{Proceedings of the 54th ACM Technical
  Symposium on Computer Science Education V. 2}}, \bibinfo{series}{SIGCSE
  2023}, \bibinfo{publisher}{Association for Computing Machinery},
  \bibinfo{address}{New York, NY, USA}, p. \bibinfo{pages}{1259},
  \doi{10.1145/3545947.3573228}.

\bibitemdeclare{book}{fitch1952symbolic}
\bibitem{fitch1952symbolic}
\bibinfo{author}{Frederic~Brenton \surnamestart Fitch\surnameend}
  (\bibinfo{year}{1952}): \emph{\bibinfo{title}{Symbolic Logic: An
  Introduction}}.
\newblock \bibinfo{publisher}{Ronald Press Company}.

\bibitemdeclare{inproceedings}{FredeKnobelsdorf2018}
\bibitem{FredeKnobelsdorf2018}
\bibinfo{author}{Christiane \surnamestart Frede\surnameend} \&
  \bibinfo{author}{Maria \surnamestart Knobelsdorf\surnameend}
  (\bibinfo{year}{2018}): \emph{\bibinfo{title}{{Explorative Datenanalyse der
  Studierendenperformance in der Theoretischen Informatik}}}.
\newblock In \bibinfo{editor}{N.~\surnamestart Bergner\surnameend},
  \bibinfo{editor}{R.~\surnamestart Röpke\surnameend},
  \bibinfo{editor}{U.~\surnamestart Schroeder\surnameend} \&
  \bibinfo{editor}{D.~\surnamestart Krömker\surnameend}, editors: {\sl
  \bibinfo{booktitle}{Hochschuldidaktik der Informatik - {HDI} 2018 - 8.
  Fachtagung des GI-Fachbereichs und Ausbilding/Didaktik der Informatik,
  Frankfurt, Germany, September 12-13, 2018}},
  \bibinfo{publisher}{Universitätsverlag Potsdam}, pp.
  \bibinfo{pages}{135--150}.
\newblock \urlprefix\url{http://eprints.cs.univie.ac.at/6870/}.

\bibitemdeclare{article}{GallegoArias2017}
\bibitem{GallegoArias2017}
\bibinfo{author}{Emilio~Jesús \surnamestart Gallego~Arias\surnameend},
  \bibinfo{author}{Benoît \surnamestart Pin\surnameend} \&
  \bibinfo{author}{Pierre \surnamestart Jouvelot\surnameend}
  (\bibinfo{year}{2017}): \emph{\bibinfo{title}{{jsCoq: Towards Hybrid Theorem
  Proving Interfaces}}}.
\newblock {\sl \bibinfo{journal}{Electronic Proceedings in Theoretical Computer
  Science}} \bibinfo{volume}{239}, p. \bibinfo{pages}{15–27},
  \doi{10.4204/eptcs.239.2}.

\bibitemdeclare{article}{gentzen:untersuchungen1}
\bibitem{gentzen:untersuchungen1}
\bibinfo{author}{Gerhard \surnamestart Gentzen\surnameend}
  (\bibinfo{year}{1935}): \emph{\bibinfo{title}{{Untersuchungen \"Uber Das
  Logische Schlie{\ss}en. I.}}}
\newblock {\sl \bibinfo{journal}{Mathematische Zeitschrift}}
  \bibinfo{volume}{35}, pp. \bibinfo{pages}{176--210},
  \doi{10.1007/BF01201353}.

\bibitemdeclare{inproceedings}{Heinze2003}
\bibitem{Heinze2003}
\bibinfo{author}{Aiso \surnamestart Heinze\surnameend} \&
  \bibinfo{author}{Kristina \surnamestart Reiss\surnameend}
  (\bibinfo{year}{2003}): \emph{\bibinfo{title}{{Reasoning and Proof:
  Methodological Knowledge as a Component of Proof Competence}}}.
\newblock In \bibinfo{editor}{Maria~Alessandra \surnamestart
  Mariotti\surnameend}, editor: {\sl \bibinfo{booktitle}{Proceedings of the
  Third Conference of the European Society for Research in Mathematics
  Education}}, pp. \bibinfo{pages}{1--10}.
\newblock
  \urlprefix\url{http://www.lettredelapreuve.org/OldPreuve/CERME3Papers/Heinze-paper1.pdf}.
\newblock \bibinfo{note}{Thematic Working Group 4, paper 5}.

\bibitemdeclare{article}{ifenthaler2016learning}
\bibitem{ifenthaler2016learning}
\bibinfo{author}{Dirk \surnamestart Ifenthaler\surnameend} \&
  \bibinfo{author}{Clara \surnamestart Schumacher\surnameend}
  (\bibinfo{year}{2016}): \emph{\bibinfo{title}{{Learning analytics im
  Hochschulkontext}}}.
\newblock {\sl \bibinfo{journal}{WiSt--Wirtschaftswissenschaftliches Studium}}
  \bibinfo{volume}{45}(\bibinfo{number}{4}), pp. \bibinfo{pages}{176--181},
  \doi{10.15358/0340-1650-2016-4-176}.

\bibitemdeclare{inproceedings}{proofbuddy2023}
\bibitem{proofbuddy2023}
\bibinfo{author}{Nadine \surnamestart Karsten\surnameend},
  \bibinfo{author}{Frederik~Krogsdal \surnamestart Jacobsen\surnameend},
  \bibinfo{author}{Kim~Jana \surnamestart Eiken\surnameend},
  \bibinfo{author}{Uwe \surnamestart Nestmann\surnameend} \&
  \bibinfo{author}{Jørgen \surnamestart Villadsen\surnameend}
  (\bibinfo{year}{2023}): \emph{\bibinfo{title}{{ProofBuddy: A Proof Assistant
  for Learning and Monitoring}}}.
\newblock In \bibinfo{editor}{Elena \surnamestart Machkasova\surnameend},
  editor: {\sl \bibinfo{booktitle}{Proceedings of the Twelfth International
  Workshop on Trends in Functional Programming in Education (TFPIE) 2023}},
  {\sl \bibinfo{series}{Electronic Proceedings in Theoretical Computer
  Science}} \bibinfo{volume}{382}, \bibinfo{publisher}{Open Publishing
  Association}, p. \bibinfo{pages}{1–21}, \doi{10.4204/eptcs.382.1}.

\bibitemdeclare{unpublished}{karsten.etal-TFPIE23}
\bibitem{karsten.etal-TFPIE23}
\bibinfo{author}{Nadine \surnamestart Karsten\surnameend},
  \bibinfo{author}{Frederik~Krogsdal \surnamestart Jacobsen\surnameend},
  \bibinfo{author}{Uwe \surnamestart Nestmann\surnameend} \&
  \bibinfo{author}{Jørgen \surnamestart Villadsen\surnameend}
  (\bibinfo{year}{2023}): \emph{\bibinfo{title}{{\textsc{ProofBuddy} ---
  Acquiring Proof Competence with Friendly Assistance}}}.
\newblock
  \urlprefix\url{https://wiki.tfpie.science.ru.nl/images/9/94/ProofBuddyDraft.pdf}.
\newblock \bibinfo{note}{Online publication of accepted pre-submissions to
  TFPIE 2023 (accessed on 2025-02-24)}.

\bibitemdeclare{inproceedings}{Kiehn2017}
\bibitem{Kiehn2017}
\bibinfo{author}{Felix \surnamestart Kiehn\surnameend},
  \bibinfo{author}{Christiane \surnamestart Frede\surnameend} \&
  \bibinfo{author}{Maria \surnamestart Knobelsdorf\surnameend}
  (\bibinfo{year}{2017}): \emph{\bibinfo{title}{{Was macht Theoretische
  Informatik so schwierig? Ergebnisse einer qualitativen Einzelfallstudie}}}.
\newblock In \bibinfo{editor}{Maximilian \surnamestart Eibl\surnameend} \&
  \bibinfo{editor}{Martin \surnamestart Gaedke\surnameend}, editors: {\sl
  \bibinfo{booktitle}{INFORMATIK 2017}}, \bibinfo{publisher}{Gesellschaft für
  Informatik, Bonn}, pp. \bibinfo{pages}{267--278}, \doi{10.18420/in2017\_20}.

\bibitemdeclare{inproceedings}{analysingStudentPracKnobelsdorf16}
\bibitem{analysingStudentPracKnobelsdorf16}
\bibinfo{author}{Maria \surnamestart Knobelsdorf\surnameend} \&
  \bibinfo{author}{Christiane \surnamestart Frede\surnameend}
  (\bibinfo{year}{2016}): \emph{\bibinfo{title}{{Analyzing Student Practices in
  Theory of Computation in Light of Distributed Cognition Theory}}}.
\newblock In: {\sl \bibinfo{booktitle}{Proceedings of the 2016 ACM Conference
  on International Computing Education Research}}, \bibinfo{series}{ICER '16},
  \bibinfo{publisher}{Association for Computing Machinery},
  \bibinfo{address}{New York, NY, USA}, pp. \bibinfo{pages}{73–--81},
  \doi{10.1145/2960310.2960331}.

\bibitemdeclare{inproceedings}{KnobKreitz2017}
\bibitem{KnobKreitz2017}
\bibinfo{author}{Maria \surnamestart Knobelsdorf\surnameend},
  \bibinfo{author}{Christiane \surnamestart Frede\surnameend},
  \bibinfo{author}{Sebastian \surnamestart B\"{o}hne\surnameend} \&
  \bibinfo{author}{Christoph \surnamestart Kreitz\surnameend}
  (\bibinfo{year}{2017}): \emph{\bibinfo{title}{{Theorem Provers as a Learning
  Tool in Theory of Computation}}}.
\newblock In: {\sl \bibinfo{booktitle}{Proceedings of the 2017 ACM Conference
  on International Computing Education Research}}, \bibinfo{series}{ICER '17},
  \bibinfo{publisher}{Association for Computing Machinery},
  \bibinfo{address}{New York, NY, USA}, p. \bibinfo{pages}{83–92},
  \doi{10.1145/3105726.3106184}.

\bibitemdeclare{inproceedings}{krieger2015mathematische}
\bibitem{krieger2015mathematische}
\bibinfo{author}{Miriam \surnamestart Krieger\surnameend} \&
  \bibinfo{author}{Kathrin \surnamestart Winter\surnameend}
  (\bibinfo{year}{2015}): \emph{\bibinfo{title}{{Mathematische
  Beweiskompetenzen Studierender diagnostizieren und fördern --- eine
  Bestandsaufnahme}}}.
\newblock In \bibinfo{editor}{Franco \surnamestart Caluori\surnameend},
  \bibinfo{editor}{Helmut \surnamestart Linneweber-Lammerskitten\surnameend} \&
  \bibinfo{editor}{Christine \surnamestart Streit\surnameend}, editors: {\sl
  \bibinfo{booktitle}{Beiträge zum Mathematikunterricht 2015, 49. Jahrestagung
  der Gesellschaft für Didaktik der Mathematik}},
  \bibinfo{organization}{Gesellschaft für Didaktik der Mathematik},
  \bibinfo{publisher}{WTM, Verlag für wissenschaftliche Texte und Medien}, pp.
  \bibinfo{pages}{508--511}, \doi{10.17877/DE290R-16693}.

\bibitemdeclare{inproceedings}{lukarov2014data}
\bibitem{lukarov2014data}
\bibinfo{author}{Vlatko \surnamestart Lukarov\surnameend},
  \bibinfo{author}{Mohamed~Amine \surnamestart Chatti\surnameend},
  \bibinfo{author}{Hendrik \surnamestart Th{\"u}s\surnameend},
  \bibinfo{author}{Fatemeh~Salehian \surnamestart Kia\surnameend},
  \bibinfo{author}{Arham \surnamestart Muslim\surnameend},
  \bibinfo{author}{Christoph \surnamestart Greven\surnameend} \&
  \bibinfo{author}{Ulrik \surnamestart Schroeder\surnameend}
  (\bibinfo{year}{2014}): \emph{\bibinfo{title}{{Data Models in Learning
  Analytics.}}}
\newblock In: {\sl \bibinfo{booktitle}{DeLFI Workshops}},
  \bibinfo{volume}{1014}, pp. \bibinfo{pages}{88--95}.
\newblock \urlprefix\url{https://ceur-ws.org/Vol-1227/paper22.pdf}.

\bibitemdeclare{misc}{TFPIE2023}
\bibitem{TFPIE2023}
\bibinfo{author}{Elena \surnamestart Machkasova\surnameend}
  (\bibinfo{year}{2023}): \emph{\bibinfo{title}{Twelfth International Workshop
  on Trends in Functional Programming in Education (TFPIE)}}.
\newblock \urlprefix\url{https://wiki.tfpie.science.ru.nl/TFPIE2023}.

\bibitemdeclare{book}{Mayring2014}
\bibitem{Mayring2014}
\bibinfo{author}{Philipp \surnamestart Mayring\surnameend}
  (\bibinfo{year}{2014}): \emph{\bibinfo{title}{Qualitative content analysis:
  theoretical foundation, basic procedures and software solution}}.
\newblock \bibinfo{publisher}{Social Science Open Access Repository, Leibniz
  Institute for the Social Sciences}.
\newblock
  \urlprefix\url{https://nbn-resolving.org/urn:nbn:de:0168-ssoar-395173}.

\bibitemdeclare{book}{mckenney2018}
\bibitem{mckenney2018}
\bibinfo{author}{Susan \surnamestart McKenney\surnameend} \&
  \bibinfo{author}{Thomas \surnamestart Reeves\surnameend}
  (\bibinfo{year}{2018}): \emph{\bibinfo{title}{Conducting educational design
  research}}.
\newblock \bibinfo{publisher}{Routledge}, \doi{10.4324/9781315105642}.

\bibitemdeclare{inproceedings}{Lean}
\bibitem{Lean}
\bibinfo{author}{Leonardo \surnamestart de~Moura\surnameend},
  \bibinfo{author}{Soonho \surnamestart Kong\surnameend},
  \bibinfo{author}{Jeremy \surnamestart Avigad\surnameend},
  \bibinfo{author}{Floris \surnamestart Doorn\surnameend} \&
  \bibinfo{author}{Jakob \surnamestart Raumer\surnameend}
  (\bibinfo{year}{2015}): \emph{\bibinfo{title}{{The Lean Theorem Prover
  (System Description)}}}.
\newblock In: {\sl \bibinfo{booktitle}{Automated Deduction --- CADE-25}}, {\sl
  \bibinfo{series}{Lecture Notes in Computer Science}} \bibinfo{volume}{9195},
  pp. \bibinfo{pages}{378--388}, \doi{10.1007/978-3-319-21401-6\_26}.

\bibitemdeclare{incollection}{nieveen2006}
\bibitem{nieveen2006}
\bibinfo{author}{Nienke \surnamestart Nieveen\surnameend},
  \bibinfo{author}{Susan \surnamestart McKenney\surnameend} \&
  \bibinfo{author}{Jan \surnamestart van~den Akker\surnameend}
  (\bibinfo{year}{2006}): \emph{\bibinfo{title}{Educational design research:
  the value of variety}}.
\newblock In: {\sl \bibinfo{booktitle}{Educational design research}},
  \bibinfo{publisher}{Routledge}, pp. \bibinfo{pages}{163--170},
  \doi{10.4324/9780203088364-21}.

\bibitemdeclare{inproceedings}{LSDproofsSemantic}
\bibitem{LSDproofsSemantic}
\bibinfo{author}{Tobias \surnamestart Nipkow\surnameend}
  (\bibinfo{year}{2012}): \emph{\bibinfo{title}{{Teaching Semantics with a
  Proof Assistant: No more {LSD} Trip Proofs}}}.
\newblock In \bibinfo{editor}{V.~\surnamestart Kuncak\surnameend} \&
  \bibinfo{editor}{A.~\surnamestart Rybalchenko\surnameend}, editors: {\sl
  \bibinfo{booktitle}{Verification, Model Checking, and Abstract Interpretation
  (VMCAI 2012)}}, {\sl \bibinfo{series}{Lecture Notes in Computer Science}}
  \bibinfo{volume}{7148}, \bibinfo{publisher}{Springer}, pp.
  \bibinfo{pages}{24--38}, \doi{10.1007/978-3-642-27940-9\_3}.

\bibitemdeclare{inproceedings}{Nipkow-CPP21}
\bibitem{Nipkow-CPP21}
\bibinfo{author}{Tobias \surnamestart Nipkow\surnameend}
  (\bibinfo{year}{2021}): \emph{\bibinfo{title}{{Teaching Algorithms and Data
  Structures with a Proof Assistant (Invited Talk)}}}.
\newblock In \bibinfo{editor}{C.~\surnamestart Hritcu\surnameend} \&
  \bibinfo{editor}{A.~\surnamestart Popescu\surnameend}, editors: {\sl
  \bibinfo{booktitle}{Certified Programs and Proofs, {CPP} 2021}},
  \bibinfo{publisher}{ACM}, pp. \bibinfo{pages}{1--3},
  \doi{10.1145/3437992.3439910}.

\bibitemdeclare{misc}{IsabelleProving}
\bibitem{IsabelleProving}
\bibinfo{author}{Tobias \surnamestart Nipkow\surnameend}
  (\bibinfo{year}{2024}): \emph{\bibinfo{title}{{Programming and Proving in
  Isabelle/HOL}}}.
\newblock \urlprefix\url{https://isabelle.in.tum.de/doc/prog-prove.pdf}.
\newblock \bibinfo{note}{Accessed on 2025-02-24}.

\bibitemdeclare{book}{Nipkow-Paulson-Wenzel:2002}
\bibitem{Nipkow-Paulson-Wenzel:2002}
\bibinfo{author}{Tobias \surnamestart Nipkow\surnameend},
  \bibinfo{author}{Lawrence~C. \surnamestart Paulson\surnameend} \&
  \bibinfo{author}{Markus \surnamestart Wenzel\surnameend}
  (\bibinfo{year}{2002}): \emph{\bibinfo{title}{{Isabelle/HOL --- A Proof
  Assistant for Higher-Order Logic}}}.
\newblock {\sl \bibinfo{series}{Lecture Notes in Computer Science}}
  \bibinfo{volume}{2283}, \bibinfo{publisher}{Springer},
  \doi{10.1007/3-540-45949-9}.

\bibitemdeclare{incollection}{plomp2013}
\bibitem{plomp2013}
\bibinfo{author}{Tjeerd \surnamestart Plomp\surnameend} (\bibinfo{year}{2013}):
  \emph{\bibinfo{title}{{Educational Design Research: An Introduction}}}.
\newblock In \bibinfo{editor}{Tjeerd \surnamestart Plomp\surnameend} \&
  \bibinfo{editor}{Nienke \surnamestart Nieveen\surnameend}, editors: {\sl
  \bibinfo{booktitle}{{Educational Design Research, Part A: An Introduction}}},
  \bibinfo{publisher}{SLO: Netherlands Institute for Curriculum Development},
  \bibinfo{address}{Enschede}, pp. \bibinfo{pages}{10--51}.
\newblock
  \urlprefix\url{https://www.slo.nl/international/@4315/educational-design/}.

\bibitemdeclare{techreport}{pop02:comparwebapplicdesktapplic}
\bibitem{pop02:comparwebapplicdesktapplic}
\bibinfo{author}{Paul \surnamestart Pop\surnameend} (\bibinfo{year}{2002}):
  \emph{\bibinfo{title}{{Comparing Web Applications with Desktop Applications:
  An Empirical Study}}}.
\newblock \bibinfo{type}{Technical Report}, \bibinfo{institution}{Link\"oping
  University}.
\newblock
  \urlprefix\url{https://www.ida.liu.se/labs/eslab/publications/pap/db/hci.pdf}.

\bibitemdeclare{inproceedings}{edukera2016}
\bibitem{edukera2016}
\bibinfo{author}{Benoit \surnamestart Rognier\surnameend} \&
  \bibinfo{author}{Guillaume \surnamestart Duhamel\surnameend}
  (\bibinfo{year}{2016}): \emph{\bibinfo{title}{{Pr{\'e}sentation de la
  plateforme Edukera}}}.
\newblock In: {\sl \bibinfo{booktitle}{Vingt-septi{\`e}mes Journ{\'e}es
  Francophones des Langages Applicatifs (JFLA 2016)}}, pp.
  \bibinfo{pages}{97--100}.
\newblock \urlprefix\url{https://hal.science/hal-01333606v1}.

\bibitemdeclare{article}{compuMethaphysics}
\bibitem{compuMethaphysics}
\bibinfo{author}{Alexander \surnamestart Steen\surnameend},
  \bibinfo{author}{Max \surnamestart Wisniewski\surnameend} \&
  \bibinfo{author}{Christoph \surnamestart Benzmüller\surnameend}
  (\bibinfo{year}{2016}): \emph{\bibinfo{title}{{Einsatz von Theorembeweisern
  in der Lehre}}}.
\newblock {\sl \bibinfo{journal}{Commentarii informaticae didacticae}}
  \bibinfo{volume}{10}, pp. \bibinfo{pages}{81--92}.
\newblock
  \urlprefix\url{https://publishup.uni-potsdam.de/opus4-ubp/frontdoor/deliver/index/docId/9485/file/cid10_S81-92.pdf}.

\bibitemdeclare{article}{tatsisa08}
\bibitem{tatsisa08}
\bibinfo{author}{Konstantinos \surnamestart Tatsisa\surnameend} \&
  \bibinfo{author}{Eugenia \surnamestart Koleza\surnameend}
  (\bibinfo{year}{2008}): \emph{\bibinfo{title}{Social and socio-mathematical
  norms in collaborative problem-solving}}.
\newblock {\sl \bibinfo{journal}{European Journal of Teacher Education}}
  \bibinfo{volume}{31}(\bibinfo{number}{1}), pp. \bibinfo{pages}{89--100},
  \doi{10.1080/02619760701845057}.

\bibitemdeclare{inproceedings}{wemmenhove2023waterproof}
\bibitem{wemmenhove2023waterproof}
\bibinfo{author}{Jelle \surnamestart Wemmenhove\surnameend},
  \bibinfo{author}{Dick \surnamestart Arends\surnameend},
  \bibinfo{author}{Thijs \surnamestart Beurskens\surnameend},
  \bibinfo{author}{Maitreyee \surnamestart Bhaid\surnameend},
  \bibinfo{author}{Sean \surnamestart McCarren\surnameend},
  \bibinfo{author}{Jan \surnamestart Moraal\surnameend},
  \bibinfo{author}{Diego~Rivera \surnamestart Garrido\surnameend},
  \bibinfo{author}{David \surnamestart Tuin\surnameend},
  \bibinfo{author}{Malcolm \surnamestart Vassallo\surnameend},
  \bibinfo{author}{Pieter \surnamestart Wils\surnameend} \&
  \bibinfo{author}{Jim \surnamestart Portegies\surnameend}
  (\bibinfo{year}{2023}): \emph{\bibinfo{title}{{Waterproof: Educational
  Software for Learning How to Write Mathematical Proofs}}}.
\newblock In \bibinfo{editor}{Julien \surnamestart Narboux\surnameend},
  \bibinfo{editor}{Walther \surnamestart Neuper\surnameend} \&
  \bibinfo{editor}{Pedro \surnamestart Quaresma\surnameend}, editors: {\sl
  \bibinfo{booktitle}{Proceedings 12th International Workshop on Theorem
  proving components for Educational software (ThEdu)}}, {\sl
  \bibinfo{series}{Electronic Proceedings in Theoretical Computer Science}}
  \bibinfo{volume}{400}, \bibinfo{publisher}{Open Publishing Association}, pp.
  \bibinfo{pages}{96--119}, \doi{10.4204/eptcs.400.7}.

\bibitemdeclare{misc}{wemmenhove2022waterproof}
\bibitem{wemmenhove2022waterproof}
\bibinfo{author}{Jelle \surnamestart Wemmenhove\surnameend},
  \bibinfo{author}{Thijs \surnamestart Beurskens\surnameend},
  \bibinfo{author}{Sean \surnamestart McCarren\surnameend},
  \bibinfo{author}{Jan \surnamestart Moraal\surnameend}, \bibinfo{author}{David
  \surnamestart Tuin\surnameend} \& \bibinfo{author}{Jim \surnamestart
  Portegies\surnameend} (\bibinfo{year}{2022}):
  \emph{\bibinfo{title}{{Waterproof: educational software for learning how to
  write mathematical proofs}}}.

\bibitemdeclare{misc}{IsabellejEdit}
\bibitem{IsabellejEdit}
\bibinfo{author}{Makarius \surnamestart Wenzel\surnameend}
  (\bibinfo{year}{2024}): \emph{\bibinfo{title}{{Isabelle/jEdit}}}.
\newblock
  \urlprefix\url{https://isabelle.in.tum.de/dist/Isabelle2024/doc/jedit.pdf}.
\newblock \bibinfo{note}{Accessed on 2025-02-24}.

\bibitemdeclare{misc}{wenzel2014isabelle}
\bibitem{wenzel2014isabelle}
\bibinfo{author}{Makarius \surnamestart Wenzel\surnameend}
  (\bibinfo{year}{2024}): \emph{\bibinfo{title}{{The Isabelle System Manual}}}.
\newblock \urlprefix\url{https://isabelle.in.tum.de/doc/system.pdf}.
\newblock \bibinfo{note}{Accessed on 2025-02-24}.

\bibitemdeclare{misc}{isabelleIsar}
\bibitem{isabelleIsar}
\bibinfo{author}{Makarius \surnamestart Wenzel\surnameend}
  (\bibinfo{year}{2024}): \emph{\bibinfo{title}{{The Isabelle/Isar Reference
  Manual}}}.
\newblock \urlprefix\url{https://isabelle.in.tum.de/doc/isar-ref.pdf}.
\newblock \bibinfo{note}{Accessed on 2025-02-24}.

\end{thebibliography}
